\documentstyle[12pt,epsf]{article}

\newcommand{\pon} {p_1}
\newcommand{\ptw} {p_2}
\newcommand{\hon} {h_1}
\newcommand{\htw} {h_2}
\newcommand{\bsigma}{\mbox{\boldmath $\sigma$}}
\newcommand{\btau}{\mbox{\boldmath $\tau$}}
\newcommand{\bqu}{{\bf q}}
\newcommand{\br}{{\bf r}}
\newcommand{\half}{\frac{1}{2}}
\newcommand{\threej}[6]{ \left( \begin{array}{ccc}
                               #1 & #2 & #3 \\
                               #4 & #5 & #6 
                             \end{array}
                        \right) } 
\newcommand{\sixj}[6]{ \left\{ \begin{array}{ccc}
                                #1 & #2 & #3 \\
                                #4 & #5 & #6 
                               \end{array}
                        \right\} } 
\newcommand{\ninej}[9]{ \left\{ \begin{array}{ccc}
                                 #1 & #2 & #3 \\
                                 #4 & #5 & #6 \\
                                 #7 & #8 & #9 
                                \end{array}
                         \right\} } 

\begin{document}

\vspace*{-2cm}

\centerline{\large {\bf Correlated model for quasi-elastic responses}}
\centerline{\large {\bf in finite nuclear systems}} 

\vspace{.3cm}

\begin{center}
{\bf Giampaolo Co'} \\
Dipartimento di Fisica, Universit\`a di Lecce  and \\
I.N.F.N. sez. di Lecce, I-73100 Lecce, Italy \\

\vspace{.1cm}

{\bf Antonio M. Lallena} \\
Departamento de F\'{\i}sica Moderna, Universidad de Granada,\\
E-18071 Granada, Spain
\end{center}

\vspace{.2cm}

\small{
A model to calculate nuclear responses considering short-range
correlation effects is presented. 
The model is applied to the study of electromagnetic responses induced
by one-body operators. 
We calculate one- and two-nucleon emission responses and cross
sections of the $^{16}$O and $^{40}$Ca nuclei in the quasi-elastic
region, and we compare them with experimental data.}

\vspace{.3cm}

\section{Introduction}
The study of atomic nuclei has been characterized in the last few
years by the development and the application of technologies able to
deal with realistic nuclear interactions.  In the nuclear structure
jargon one calls realistic those interactions whose parameters have
been fixed to reproduce the properties of two, and eventually three,
nucleon systems.

A large number of theories has been developed to tackle the problem of
solving the Schr\"odinger equation with realistic interactions.  Three
or four nucleon systems are described with Faddeev \cite{che86,sta91},
Correlated Hyperspherical Harmonics Expansion \cite{kie93,efr00} and
Green Function Monte Carlo \cite{pud97} techniques which solve the
Schr\"odinger equation without approximations. The Green Function
Monte Carlo technique has also been applied, with great success, to
nuclei up to A=7 \cite{pud97}.

Mainly for computational reasons, the straightforward applications of
the above theories to heavier nuclear systems is not affordable,
therefore different approaches such as Cluster Monte Carlo
\cite{pie92}, Brueckner theory \cite{sch91}, Exponential S
\cite{hei99}, Correlated Basis Function (CBF) \cite{fab00}, have been
developed with the aim of obtaining approximate, but still accurate,
solutions of the many-body Schr\"odinger equation.

Among these theories the CBF \cite{cla79} has been well tested and
widely applied in infinite nuclear systems like nuclear and neutron
matter \cite{wir88}-\cite{akm98}.  These studies have shown that, in
the framework of the CBF theory, the Fermi Hypernetted Chain (FHNC)
resummation technique with the Single Operator Chain (SOC)
approximation provides solutions of the Schr\"odinger equation having
an accuracy of about 1 MeV per nucleon on the binding energy. In this
computational scheme the use of Argonne nucleon-nucleon potentials
\cite{wir84,wir95}, together with Urbana three-body forces
\cite{sch86}, provides equations of state whose minima are rather
close to the empirical one. In the same framework also the nuclear
matter responses to external probes have been calculated obtaining
satisfactory agreement with experimental inelastic electron scattering
data \cite{fan87,fab89}.

The success of the infinite nuclear matter results has lead one to
apply the CBF theory to the description of the ground state of finite
nuclear systems.  Recently, $^{16}$O and $^{40}$Ca ground states have
been described within the FHNC-SOC computational scheme with
interactions including tensor, spin-orbit terms and three-body forces
\cite{fab00}.

In this paper we deal with the problem of describing excited states of
finite nuclei within the CBF computational scheme.  The model we
present is inspired by the nuclear matter works of
Refs. \cite{fan87,fab89}.  While in these works the CBF cluster
expansion has been fully considered, in our model we retain only those
terms containing a single correlation function.  The validity of this
truncation has been tested, for the nuclear matter charge response, in
Ref. \cite{ama98}.  Because of the excellent agreement between the
full calculation and our truncated model we felt confident enough to
extend our model to include the current operators.

Our model treats the short-range correlations but does not consider
collective nuclear excitations. For this reason we have applied it to
the description of the nuclear responses in the quasi-elastic region
where the collective effects are negligible \cite{smi88,co88}.

In this paper we shall deal only with inclusive electron scattering
data, and we consider excited states with both one and two
particles in the continuum. The two nucleon emission is treated as a
genuine short-range correlation effect, and we neglect
the contribution of the two-body currents.

After briefly reviewing in section \ref{sect:response}
the approach of Ref. \cite{fan87}, we present in section \ref{sect:model}
our model, and we apply it in section \ref{sect:emexcitations}
to inclusive electron scattering. 
The results obtained in the calculations of the $^{16}$O and $^{40}$Ca
quasi-elastic responses are presented in section \ref{sect:sa}.
In section \ref{sect:concl} we summarize our work and 
we draw our conclusions.

\section{Responses in a correlated theory}
\label{sect:response}
The linear response of a many-body system to the perturbations induced
by an external operator $O(\bqu)$ is given by:
\begin{equation}
S(\bqu,\omega) \, =\, - \frac{1}{\pi} \, \rm{Im} \, D(\bqu,\omega) \, ,
\label{nresp}
\end{equation}
with
\begin{equation}
D(\bqu ,\omega) \,= \,
\langle \stackrel{\large \sim}{\Psi}_0 |\,O^{+}(\bqu)\,
(H - E_0 - \omega + i \eta)^{-1}\,O (\bqu) \,
|\stackrel{\sim}{\Psi}_0 \rangle  \, ,
\label{d1}
\end{equation}
where we have indicated with
$|\stackrel{\sim}{\Psi}_n \rangle$ the normalized eigenstates of
the nuclear hamiltonian $H$:
\begin{equation}
|\stackrel{\sim}{\Psi}_n \rangle \, = \,
\frac {|\Psi_n \rangle } {\langle \Psi_n |\Psi_n\rangle ^{\half}} \, .    
\label{psinorm}
\end{equation}
In the previous equations $\bqu$ and $\omega$ represent  
the momentum and energy transferred to the nucleus.

Inserting in Eq. (\ref{d1}) a complete set of eigenvectors of $H$ we
obtain:
\begin{equation}
D(\bqu ,\omega) \,= \,\sum_n \,\frac { |\langle \stackrel{\sim}{\Psi}_n |\,
 O(\bqu)\, |\stackrel{\sim}{\Psi}_0 \rangle |^2 } 
{E_n - E_0 - \omega + i \eta }\, =\,
 \sum_n \, \xi_n^+(\bqu)\, (E_n - E_0 - \omega + i \eta)^{-1} \, \xi_n(\bqu)
\, ,
\label{d2}
\end{equation}
where we have defined:
\begin{equation}
\xi_n(\bqu)\, =\, \frac { \langle \Psi_n | \,O(\bqu)\, |\Psi_0 \rangle }
 {\langle \Psi_n |\Psi_n\rangle ^{\half}\,\langle \Psi_0 |\Psi_0\rangle
 ^{\half} }\, .
\label{xi1}
\end{equation}

Like in CBF theory we assume that the nuclear many-body ground
state is described as a product
of a correlation function $G$ and a Slater determinant
$|\Phi_0\rangle$ of a set of single particle wave functions occupying
all, and only, the states below the Fermi surface:
\begin{equation}
|\Psi_0\rangle \,= \,G \,|\Phi_0\rangle \, .
\label{psi0def}
\end{equation}
Since the Slater determinant is already antisymmetrized with respect
to the exchange of two nucleons, the correlation operator is given by
a symmetrized product of two-body correlation operators:
\begin{equation}
G(1,2...A)\,=\,{\cal S}\, \left[\, \prod_{i<j}\, F_{ij}\, \right] \, ,
\label{cordef}
\end{equation}
where we have indicated with $\cal S$ the symmetrizer operator.

In modern nuclear structure calculations with realistic microscopic
interactions, the two-body correlation operator is taken
as a sum of operator dependent correlation functions
\begin{equation}
F_{ij}\,=\, \sum_{p=1,8}\, f^p(r_{ij})\, O^p_{ij} \, ,
\label{cortwo}
\end{equation}
where the involved operators are:
\begin{equation}
O^{p=1,8}_{ij}\,=\,
\left[ 1, \, \bsigma_i \cdot \bsigma_j, \, S_{ij}, \,
({\bf L} \cdot {\bf S})_{ij} \right]\, \otimes \,
\left[ 1,\,  \btau_i \cdot \btau_j \right] \, ,
\label{oper}
\end{equation}
with 
$S_{ij}=(3\,{\hat  {\bf r} }_{ij} \cdot \bsigma_i  \,
{\hat{\bf r}}_{ij} \cdot 
\bsigma_j -  \bsigma_i \cdot \bsigma_j)$ 
indicating the tensor operator, and  $r_{ij}=|{\bf r}_i-{\bf r}_j|$
the distance between the positions of the particles $i$ and $j$.

The ground state wave function is obtained by minimizing the ground
state energy with respect to variations of the correlation function
and of the single particle basis. 
In the theory developed in Ref. \cite{fan87}, the 
correlation operator $G$ whose parameters have been fixed by the
ground state minimization is used to generate the excited states of
the system:
\begin{equation}
|\Psi_n\rangle \,= \,G \,|\Phi_n\rangle \, .
\label{psidef}
\end{equation}
The mean-field excited states $|\Phi_n\rangle$ are obtained by making
particle-hole excitations on $|\Phi_0\rangle$.

In order to use the cluster expansion techniques it is useful to rewrite
the function $\xi_n(\bqu)$ as:
\begin{equation}
\xi_n(\bqu) \,= \,\frac {\langle \Phi_n |\,G^+ \,O(\bqu) \,G|\Phi_0 \rangle
}{\langle \Phi_0|\,G^+\,G\,|\Phi_0\rangle } \left[ \frac {\langle \Phi_0
|\,G^+\,G\,|\Phi_0\rangle }{\langle \Phi_n|\,G^+\,G\,|\Phi_n\rangle }
\right]^\half \, .
\label{xi2}
\end{equation}
The two factors in Eq. (\ref{xi2}) are evaluated separately by
expanding the numerator and the denominator in powers of the
short-range function $F_{ij} -1$ (see Ref. \cite{fan87} for a detailed
presentation of the cluster expansion of $\xi_n(\bqu)$).  In both
factors the denominators cancel the unlinked diagrams.

\section{The model}
\label{sect:model}
Our model simplifies the calculation of the cluster expansion of 
Eq. (\ref{xi2}) by retaining only those terms involving a single
correlation line. This model has been
inspired by the results of Refs. \cite{co95,ari97}
where the density and momentum distributions of doubly closed shell
nuclei, calculated with a single correlation line
model, have been shown to be rather similar to those obtained with
complete CBF/FHNC calculations.

In the present article we consider that the external operator
$O(\bqu)$ is a one-body operator and
we use only scalar correlations. Considering
$O^p_{ij}=0$ for $p>1$ in Eq. (\ref{oper}), we can express
the function $\xi_n(\bqu)$ as:
\begin{eqnarray}
\nonumber \xi_n(\bqu) &=& 
\frac {\langle \Phi_n | \prod_{i<j}\,f^+_{ij}\, O(\bqu)
\,\prod_{i<j}\,f_{ij} \, |\Phi_0 \rangle } 
{\langle \Phi_0|\prod_{i<j}\,f^+_{ij} \,\prod_{i<j}\,f_{ij}\, |\Phi_0\rangle } 
\,\left[ \frac {\langle \Phi_0 |\prod_{i<j}\,f^+_{ij} \,\prod_{i<j}\,f_{ij}
\, |\Phi_0\rangle } {\langle \Phi_n|\prod_{i<j}\,f^+_{ij}\,
\prod_{i<j}\,f_{ij}\, |\Phi_n\rangle } \right]^\half \\ &=& \frac {\langle
\Phi_n |\,O(\bqu)\, \prod_{i<j}\,(1+h_{ij}) \,|\Phi_0 \rangle } {\langle
\Phi_0|\prod_{i<j}\,(1+h_{ij}) \,|\Phi_0\rangle }\, \left[ \frac {\langle
\Phi_0 | \prod_{i<j}\,(1+h_{ij}) \,|\Phi_0\rangle }
{\langle \Phi_n|\prod_{i<j}\,(1+h_{ij}) \, |\Phi_n\rangle } 
\right]^\half \, ,
\label{xi3}
\end{eqnarray}
where we have considered that $f_{ij}$ is a real scalar function, 
therefore commuting with the operator $O(\bqu)$, and we
have defined $h_{ij}=f^2_{ij}-1$.

The approximation of our model consists in retaining in Eq.(\ref{xi3})
only those diagrams having a single $h$ function:
\begin{equation}
\xi_n(\bqu) \rightarrow \xi^1_n(\bqu) \,=\,
\langle \Phi_n |\,O(\bqu)\, \sum_{i<j}\, (1+h_{ij}) \,|\Phi_0 \rangle _L \, ,
\label{ximodel}
\end{equation}
where the subindex $L$ indicates that only the linked diagrams are
considered. 
This result has been obtained using the same procedure used in
Ref. \cite{co95} for the density distribution. 
The first step consists in making the full cluster expansion of 
numerators and denominators, and this allows the elimination of 
the unlinked diagrams. Only at this point do we truncate the obtained 
result by considering only the first order terms in $h_{ij}$.

\begin{figure}
\begin{center}
\hspace*{-2.0cm}
\leavevmode
\epsfysize = 300pt
\epsfbox[70 200 500 650]{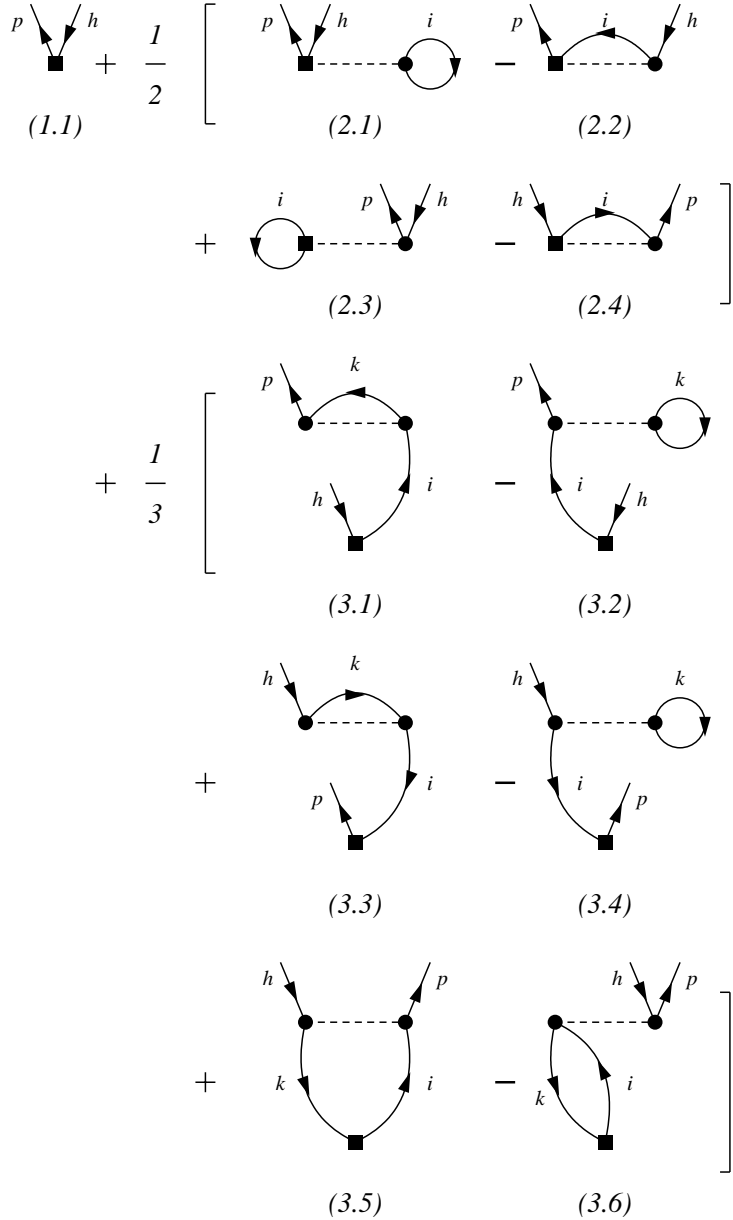}
\end{center}
\vspace{2.5cm}
\caption{\small Diagrams considered in the calculation of the
one-particle one-hole responses. The dashed lines represent the
correlation function $h$ and the continuous oriented lines the single
particle wave functions. The letter $i,k,h$ indicates hole wave
functions and $p$ particle wave functions. A sum on the $i$ and $k$
indexes is understood. The black squares represent the point where the
excitation operator is acting. }
\label{fig:diag1p1h}
\end{figure}

\begin{figure}
\begin{center}
\hspace*{-2.0 cm}
\leavevmode
\epsfysize = 300pt
\epsfbox[70 200 500 650]{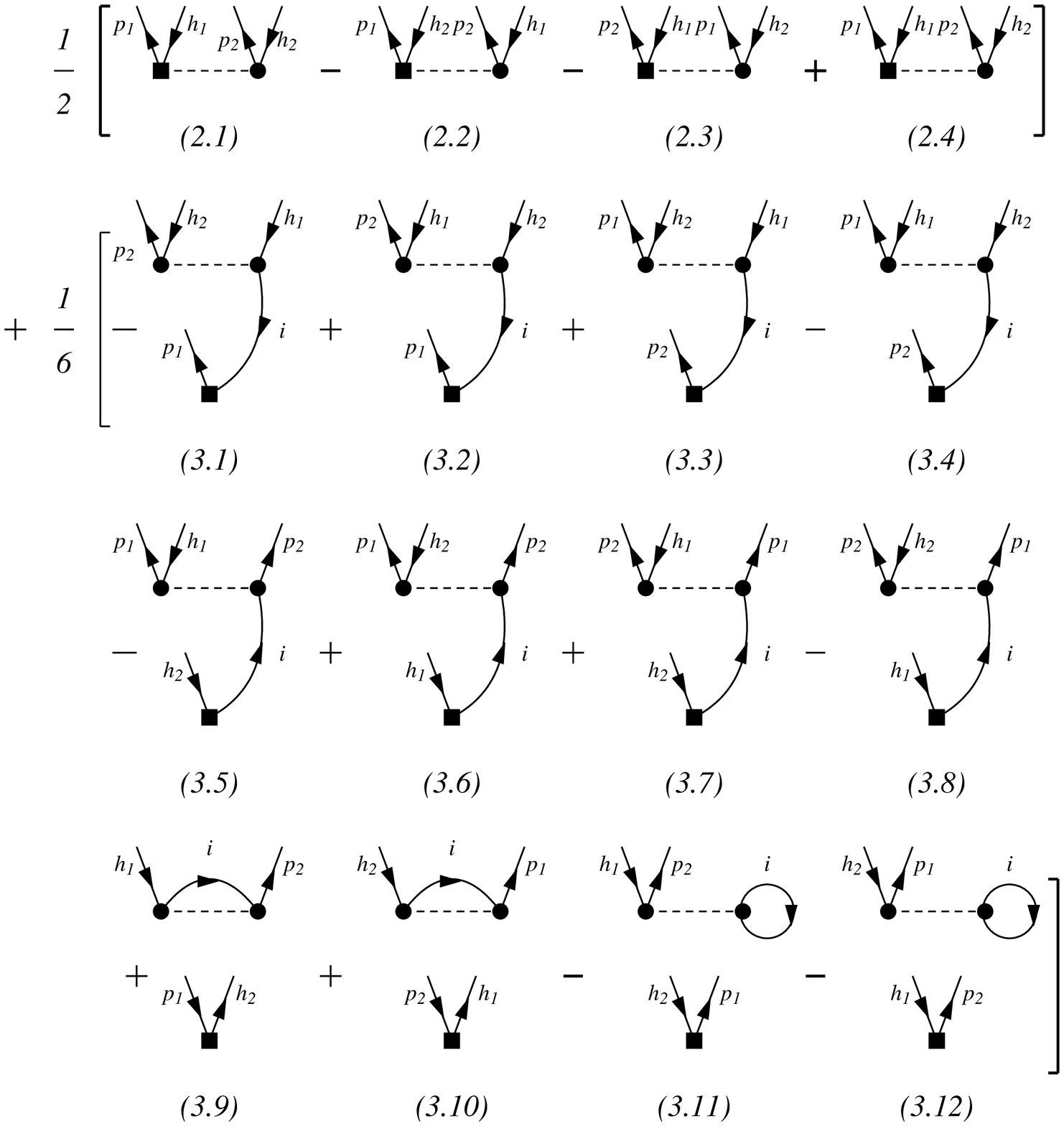}
\end{center}
\vspace{.5cm}
\caption{\small Diagrams considered in the calculation of the
two-particle two-hole responses. The symbols have the same meaning as
in Fig. ~\protect\ref{fig:diag1p1h}. }
\label{fig:diag2p2h}
\end{figure}

The terms contributing to $\xi^1_n(\bqu)$ are presented as
Mayer-like diagrams in Figs. 1 and 2 for 1p-1h and 2p-2h excitations 
respectively. In each diagram, the black square indicates the 
coordinate where the excitation one-body operator $O(\bqu)$ is acting, 
while the black dots indicate the other coordinates.
The dashed line indicates the correlation function $h$, which operates
on two-coordinates only, and the continuous oriented lines indicate the
single particle wave functions $\phi$. 
We have used the convention of considering
entering into a point the wave functions of the state
$|\Phi\rangle $, while the
wave functions of the state $\langle \Phi|$ are exiting from the point.
The letters $i,j,k$ indicate single particle wave functions below the
Fermi surface and imply a summation over them. With $p$ and $h$ we
have labelled those particle and hole states whose quantum numbers
characterize the full many-body excited state.

Let us first consider the case when the nuclear final states have only
one particle in the continuum.
If we label with $1$ the coordinate where the external
operator $O(\bqu)$ is acting, 
we can specify the $\xi ^1$ in Eq. (\ref{ximodel}) as:
\begin{eqnarray}
\nonumber
\xi^1_{\rm 1p1h}(\bqu) \,  &=&\,
\langle \Phi_{\rm 1p1h} |\, O(\bqu)\, |\Phi_0 \rangle  
\, + \, \langle \Phi_{\rm 1p1h} |\, O(\bqu) \, 
\sum^{A}_{j>1}\, h_{1j}\, |\Phi_0 \rangle \\
\, &~& + \, \langle \Phi_{\rm 1p1h} |\, O(\bqu) \, \sum^{A}_{1< i <j } \,
h_{ij} \, |\Phi_0 \rangle \, .
\label{xi1p1h}
\end{eqnarray}

The above expression shows that our model, in addition to the 
uncorrelated transitions represented by the
the one-point diagram (1.1) in
Fig. 1, generates also two- and three-point
diagrams. The presence of both two- and three-point diagrams is
necessary to have a correct normalization of the many body wave
function.
This can be seen by considering that, if $O(\bqu)$ is the density
operator, in the limit for $q \rightarrow 0$, when $p = h$, the sum
on all the hole single particle wave functions 
should provide the proper number of nucleons $A$, as
happens in Refs. \cite{co95,ari97}. In this limit, 
the diagrams containing a correlation line should not
contribute. This can be seen
in Fig. 1, by joining the $p$ and $h$ lines
and considering that in the three-point diagrams 
the $q \rightarrow 0$ limit implies that the wave functions linked to
the black square should be equal because of
the orthonormality of the set of single particle wave functions.
One can then observe that the contribution of the (2.1) and (2.3) diagrams
is exactly canceled by that of the (3.2), (3.4) and (3.6) diagrams,
and the contribution of the (2.2) and (2.4) diagrams
is canceled by that of the (3.1), (3.3) and (3.5) diagrams.

When the nuclear final state is
characterized by two-particle two-hole excitations,
the function $\xi_n^1(\bqu)$ is given by:
\begin{equation}
\nonumber
\xi^1_{\rm 2p2h}(\bqu) \,=\,
\langle \Phi_{\rm 2p2h} |\,O(\bqu) \,\sum^{A}_{1<j}\,h_{1j}\, |\Phi_0 \rangle 
\,+\, \langle \Phi_{\rm 2p2h} |O(\,\bqu) \,\sum^{A}_{1< i  <j }
\,h_{ij}\, |\Phi_0 \rangle \, .
\label{xi2p2h}
\end{equation}

As expected,
the uncorrelated term does not appear, since a one-body operator
cannot lead to a 2p-2h final state. 
The 4 two-point diagrams and the 12 three-point
diagrams we consider are shown in Fig. 2.
Also in this case the set of diagrams conserves the proper
normalization. In analogy with the
discussion done for the 1p-1h case one
can see that, in the limit $q \rightarrow 0$ and setting 
$p_1=h_1$ and $p_2=h_2$ the
contribution of the diagrams of Fig. 2 is zero, as
expected. 

\section{Electromagnetic excitations}
\label{sect:emexcitations}
The response model presented in the previous sections
has been applied to the description of inclusive electron scattering
processes considering both one- and two- nucleon emission.

In Plane Wave Born Approximation,
the inclusive electron scattering cross section can be written as:
\begin{eqnarray}
\nonumber
\frac{{\rm d}^2 \sigma}{{\rm d} \Omega' {\rm d} \epsilon'} &=& 
\sigma_M 
\left\{ \,
\frac{q^4_\mu}{ \bqu^4} 
\left[ R^{\rm 1p1h}_L(q,\omega) + R^{\rm 2p2h}_L(q,\omega) \right]
\right. \\
 &~&  \left.  +
\left(\tan ^2 \frac{\theta}{2}- \frac{q^2_\mu}{ 2\bqu^2} 
\right)\,
\left[ R^{\rm 1p1h}_T(q,\omega) + R^{\rm 2p2h}_T(q,\omega) \right]\,
\right\} \, ,
\label{cross}
\end{eqnarray}
where $\theta$ is the scattering angle,
$q^\mu$ the four-momentum transfer satisfying the relation
$q^2_\mu=\omega^2 - \bqu^2$, and 
$\sigma_M$ is the Mott cross section:
\begin{equation}
\sigma_M = \left( \frac{\alpha \cos(\theta/2)}
                       {2 \epsilon_i \sin^2(\theta/2)} \right)^2
\, .
\end{equation}
In the last equation $\alpha$ is the fine structure constant and
$\epsilon_i$ is the initial energy of the  electron.

The electromagnetic responses $R_L$ and $R_T$,
depending on $q \equiv |\bqu|$ and $\omega$,
are obtained by evaluating the responses
$S(\bqu,\omega)$ in Eq. (\ref{nresp}) for charge and current
operators respectively, and multiplying them by the electromagnetic
nucleon form factors.
In Eq. (\ref{cross}) the 1p-1h and 2p-2h responses do not
interfere with each other because they produce two different
final states.

We have calculated the longitudinal responses using the charge
operator:
\begin{equation}
O(\bqu) \, \rightarrow \, \rho (\bqu) \, =  \,
\int {\rm d}^3 r \,\, e^{i \bqu \cdot \br} \, \rho (\br) \, ,
\label{chq}
\end{equation}
with
\begin{equation}
\rho (\br) \,= \, \sum^A_{k=1} \, \frac {1+\tau_3(k)}{2} \,
\delta(\br -\br_k) \, ,
\label{chr}
\end{equation}
where $\tau_3(k)$ is the third component of the isospin of the
$k$-th nucleon.

The transverse responses have been calculated considering
convection and  magnetization currents:
\begin{equation}
O(\bqu) \, \rightarrow \, {\bf J} (\bqu) 
={\bf j}^c (\bqu) + {\bf j}^m (\bqu)
= \, \int {\rm d}^3 r \,\, e^{i \bqu \cdot \br} 
\, [ \, {\bf j}^c (\br) + {\bf j}^m (\br) \, ] ,
\label{cuq}
\end{equation}
with
\begin{equation}
{\bf j}^c (\br) \,= \, \sum^A_{k=1} \, \frac{1}{i2M_k} \,
\frac{1+\tau_3(k)}{2} \,  \left[ \delta(\br -\br_k) \, \nabla_k \, 
+ \, \nabla_k \, \delta(\br -\br_k) \right] 
\label{conv}
\end{equation}
and
\begin{equation}
{\bf j}^m (\br) \,= \, \sum^A_{k=1} \, \frac{1}{M_k} \,
\left(\mu_P \frac{1+\tau_3(k)}{2} + \mu_N \frac {1 - \tau_3(k)}{2} \right) 
\, \nabla_k \times \delta(\br -\br_k) \, \bsigma(k) \, .
\label{cur}
\end{equation}
In the above equation we have indicated with $M_k$ and $\bsigma(k)$ 
the mass and Pauli spin matrices corresponding to the $k$-th nucleon, and
with $\mu_P$ ($\mu_N$) the proton (neutron) anomalous magnetic moment.

Since in the quasi-elastic peak the contribution of ${\bf j}^c$ is
small compared to that of ${\bf j}^m$ (see the discussion of Fig. 5 
in the next section) we have considered the convection current only 
in the uncorrelated one-body term (1.1) of Fig. 1.   

We restrict our study to the investigation of doubly closed shell
nuclei and we suppose that the target nucleus makes a transition 
from its ground state
to an excited state characterized by the total angular momentum $J$,
its projection $M$ on the quantization axis, and the parity 
$\Pi=\pm 1$. The responses can be written as:
\begin{equation}
R_L(q,\omega) \, = \, \sum^\infty_{J,M} \,
|\langle \Psi^{\Pi}_{JM}|\, M_{JM}(q)\, |\Psi^1_{00}\rangle |^2 \,
\delta (E_J - E_0 - \omega) 
\label{rljm}
\end{equation}
and
\begin{equation}
R_T(q,\omega)\, = \, \sum^\infty_{J,M}\, \left\{ |\langle
\Psi^{\Pi}_{JM}|\, T^{\rm E}_{JM}(q)\, |\Psi^1_{00}\rangle |^2 \, + \,
|\langle \Psi^{\Pi}_{JM}|\, T^{\rm M}_{JM}(q)\,|\Psi^1_{00}\rangle |^2
\right\}\, \delta (E_J - E_0 - \omega) \, ,
\label{rtjm}
\end{equation}
where we have not indicated the parity of the ground state which is
always positive in our case.
The $M_{JM}$ and $T_{JM}$
operators used in the previous equations are obtained by
making a multipole expansion of Eqs. (\ref{chq}) and
(\ref{cuq}). For the charge we have:
\begin{equation}
M_{JM}(q) \,= \, \int {\rm d}^3r \, j_J(qr) \, Y_{JM}(\hat{\br}) 
 \, \rho(\br) \, ,
\label{mjm}
\end{equation}
where we have indicated with $\hat{\br}$ the $\theta$ and $\phi$ 
angles characterizing the
vector $\br$ in polar coordinates, with $Y_{JM}$ the spherical
harmonics and  with $j_J$ the spherical Bessel functions.

For the current we should distinguish between the electric excitations (E)
with natural parity $\Pi=(-1)^J$ and the magnetic excitations (M) with
unnatural  parity $\Pi=(-1)^{J+1}$:
\begin{equation}
T^{\rm E}_{JM}(q) \, = \, \frac{1}{q} \,
\int {\rm d}^3r \, \left\{ \nabla \times 
\left[ j_J(qr) \, {\bf Y}^M_{JJ}(\hat{\br}) \right ] \right\} 
\cdot {\bf J} (\br) 
\label{tejm}
\end{equation}
and
\begin{equation}
T^{\rm M}_{JM}(q) \, = \,
\int {\rm d}^3r \, j_J(qr) \,{\bf Y}^M_{JJ}(\hat{\br}) \cdot 
{\bf J} (\br) \, .
\label{tmjm} 
\end{equation}
where we have used the symbol  ${\bf Y}^M_{JJ}$ to indicate the 
vector spherical harmonics \cite{edm57}.

We calculate the responses using the expressions (\ref{nresp}),
(\ref{d2}) and (\ref{ximodel}), and we rewrite Eqs. (\ref{rljm}) and
(\ref{rtjm}) as:
\begin{equation}
R_L(q,\omega)\, = \, 4 \pi \,
\sum^\infty_J \,
|\langle \Phi^{\Pi}_J \| \, M_J(q) \, \sum_{i<j} \, (1+h_{ij}) \, 
\|\Phi^{1}_0\rangle |^2 \,
\delta (E_J - E_0 - \omega) 
\label{rlj}
\end{equation}
and
\begin{eqnarray}
\nonumber
R_T(q,\omega) &=& 4 \pi \, \sum^\infty_J \,
\left\{|\langle \Phi^{\Pi}_J \| \, T^{\rm E}_J(q)\, \sum_{i<j}\, (1+h_{ij}) \, 
\|\Phi^{1}_0\rangle |^2 
\right. \\
&& + \left.
|\langle \Phi^{\Pi}_J \| \, T^{\rm M}_{JM}(q) \, \sum_{i<j}\, (1+h_{ij}) \, \|
\Phi^{1}_0 \rangle |^2 \right\} \, \delta (E_J - E_0 - \omega) \, .
\label{rtj}
\end{eqnarray}

The above expressions have been obtained by applying the Wigner-Eckart
theorem, therefore the symbol ``$\,\|\,$''  indicates that
the angular part should be calculated by considering only the reduced
matrix elements. 

The calculations of the transition matrix elements are carried out by
performing a multipole expansion of the correlation function $h$:
\begin{equation}
h_{ij} \,= \,h(r_{ij}) \,= \, h(r_i,r_j,\cos \theta_{ij}) \,
 =  \, \sum_{L=0}^\infty  \, h_L(r_i,r_j)  \, P_L(\cos \theta_{ij}) \,
,
\label{hexp}
\end{equation}
where $P_L$ are Legendre polynomials.
Because of the completeness and the orthogonality 
of the Legendre polynomials,
the $ h_L(r_i,r_j)$ terms can be evaluated as: 
\begin{equation}
h_L(r_i,r_j)  \, =  \, \frac {2L+1}{2} \,
\int_{-1}^{1} {\rm d}(\cos \theta_{ij})  \, P_L(\cos \theta_{ij}) \,
h(r_i,r_j,\cos \theta_{ij})   \, .
\label{hldef}
\end{equation}

Eqs. (\ref{rlj}) and (\ref{rtj}) show that the final state
quantum numbers are determined by the quantum numbers
of the uncorrelated many body state $|\Phi^\Pi_{JM}\rangle$
which is built as a Slater determinant of single particle
wave functions of the form:
\begin{equation}
\phi_k(\br) \, \equiv \,
R^t_{nlj}(r) \, 
\sum_{\mu s} \, \langle  l \mu \half s | j m \rangle  \,
 Y_{l\mu}(\hat{\br}) \, \chi_s \, \chi_t \, ,
\label{spwf}
\end{equation} 
where $\chi_s$ and $\chi_t$ are
the spin and isospin wave functions respectively
and $ \langle  l \mu \half s | j m \rangle  $ 
a Clebsch-Gordan coefficient.

We treat now separately the one and two nucleon emission cases.
In the first one, once the hole is fixed, energy conservation defines
the value of the emitted particle energy:
$\epsilon_p=\omega+\epsilon_h$. 
The angular momentum coupled Slater determinant is related to the
uncoupled one by the relation:
\begin{equation}
|\Phi^{\Pi}_{JM}({\rm 1p1h})\rangle   \,=  \,
\sum_{m_p m_h}  \,(-1)^{j_h - m_h}  
\, \langle  j_p m_p j_h -m_h | JM \rangle  \, 
|\Phi^{\Pi}_{\rm 1p1h}\rangle  \, ,
\label{phi1p1h}
\end{equation}
where $j_p$ and $j_h$ are the particle and hole angular momenta and
$m_p$ and $m_h$ their third components.
We insert the expression (\ref{phi1p1h}) into Eq.
(\ref{xi1p1h}) and  apply the Wigner-Eckart theorem. Using the
orthornormality of the single particle wave functions we obtain the 
following expression:
\begin{equation}
\xi^{1,J}_{\rm 1p1h}(\bqu) \, = \, \langle p \| \, O_J(\bqu)\, \|h\rangle
\, + \, \sum^{A}_{j} \, \langle pj \| \,O_J (\bqu)\, h_{12} \, \| hj
\rangle \, +\, \sum^{A}_{j , k }\, \langle pjk \| \,O_J(\bqu)\, h_{23}
\| hjk\rangle \, .
\label{xi1p1hj}
\end{equation}
In this equation we have labelled with 1 the coordinate where
the electromagnetic one-body operator is acting.
The second and the third terms of Eq. (\ref{xi1p1hj}) produce the
two- and three-point diagrams of Fig. 1
respectively. 

The multipole expanded expression of a generic one-body operator 
is formed by the product of a term depending on $q$ and $\br_1$
multiplied by
a term depending only on the angular coordinates:
\begin{equation}
O_{JM}(\bqu)=F_J(qr_1)O_{JM}(\hat{\br}_1) .
\label{divop}
\end{equation}

In the case we are interested this is obtained 
by inserting the expressions (\ref{chr}), (\ref{conv}) and  (\ref{cur}) 
into Eqs. (\ref{mjm}), (\ref{tejm}) and (\ref{tmjm}).

We present in the following the expressions of the terms of
Eq. (\ref{xi1p1hj}) using the notation of Ref. \cite{edm57} to indicate
the 3j, 6j and 9j symbols of Wigner.
For the one-point diagram  we obtain:
\begin{equation}
D^{(\rm 1p1h)}_1 \, = \,\omega_{ph}^J 
\int {\rm d}r_1 \, r_1^2 \, R^*_p(r_1) \,
F_J(qr_1) \, R_h(r_1) \, ,
\end{equation}
where we have defined:
\begin{equation}
\omega_{\alpha \beta}^K \, = 
\langle l_\alpha \half j_\alpha \, \| \, O_K(\hat{\br}_1)  \, \|
\, l_\beta \half j_\beta \rangle \, . 
\end{equation}

The expressions we have obtained for the two-point diagrams are:
\begin{eqnarray}
\label{D21}
D^{(\rm 1p1h)}_{2.1} & = & 4\pi \,\omega_{ph}^J 
 \int {\rm d}r_1 \, r_1^2 \, R^*_p(r_1) \,
  F_J(qr_1)
\, R_h(r_1) \, {\cal H}_0^{[\rho]}(r_1) \, , \\[5mm]
D^{(\rm 1p1h)}_{2.2} & = & 4\pi \, (-1)^{j_p+j_h+J} \,
\sum_{i L K} \, \delta_{t_i,t_h} \,
(-1)^L \, \frac{\widehat{K}}{{\widehat{L}}^2} \, \gamma_{i h}^L \, 
\sixj{L}{K}{J}{j_p}{j_h}{j_i} \\&&
\omega_{pi}^{LJ;K}\, \int {\rm d}r_1 \, r_1^2 \, R^*_p(r_1) \, 
F_J(qr_1)
R_i(r_1) \, {\cal H}_L^{[ih]}(r_1) \nonumber \, , \\[5mm]
D^{(\rm 1p1h)}_{2.3} & = & 4\pi \, \delta_{t_p,t_h} \,(-1)^J \,
\frac{1}{{\widehat{J}}^3} \, \gamma_{ph}^J \, \sum_{i} \, \widehat{j}_i
\, \omega_{ii}^{JJ;0} \\ && \nonumber
\int {\rm d}r_1 \, r_1^2 \, R^*_i(r_1) \, 
 F_J(qr_1)
R_i(r_1) \, {\cal H}_J^{[ph]}(r_1) \, , \\[5mm]
D^{(\rm 1p1h)}_{2.4} & = & 4\pi \, (-1)^{j_p+j_h} \,
\sum_{i L K} \, \delta_{t_i,t_p} \,
(-1)^K \, \frac{\widehat{K}}{{\widehat{L}}^2} \, \gamma_{pi}^L \, 
\sixj{L}{K}{J}{j_h}{j_p}{j_i} ~~~~~~~~~~~~\\&&
\omega_{ih}^{LJ;K} \, \int {\rm d}r_1 \, r_1^2 \, R^*_i(r_1) \,
 F_J(qr_1) \,
R_h(r_1) \, {\cal H}_L^{[pi]}(r_1) \nonumber \, .
\end{eqnarray}
\noindent
In the previous equations, we have used the notation 
$\widehat{\alpha}=\sqrt{2\alpha+1}$ and the following
definition of $\gamma^{\lambda}_{\alpha \beta}$:
\begin{eqnarray} 
\nonumber \gamma^{\lambda}_{\alpha \beta} & = & 
\langle l_\alpha \frac12 j_\alpha \, \| \, Y_\lambda(\hat{\bf r}) \, \|
\, l_\beta \frac12 j_\beta \rangle \\
& = & \frac{1}{\sqrt{4\pi}} \, (-1)^{j_{\beta}+\lambda-\frac12} \,
\xi(l_\alpha+l_\beta+\lambda) \, \widehat{j}_\alpha \widehat{j}_\beta
\widehat{\lambda} \,
\threej{j_\alpha}{j_\beta}{\lambda}{\half}{-\half}{0} \, ,
\end{eqnarray}
with $\xi(\alpha)$=1 if $\alpha$ is even and $\xi(\alpha)$=0 otherwise.
The other symbols used are:
\begin{equation}
\omega_{\alpha \beta}^{LJ;K} \, = \, 
\langle \alpha \| \left[Y_L (\theta_1,\phi_1)
\otimes O_J(1) \right]^K \| \beta \rangle 
\end{equation}
and the integrals
\begin{equation}
\label{calHint}
{\cal H}_L^{[\alpha \beta]}(r_1) \, = \, \int {\rm d}r_2 \, r_2^2 \,
R^*_\alpha(r_2) \, h_L(r_1,r_2) \, R_\beta(r_2) 
\end{equation}
and, in Eq.~(\ref{D21}),
\begin{equation}
\label{calH0rho}
{\cal H}_0^{[\rho]} \, = \, \frac{1}{4\pi} \,
\sum_i \, {\widehat{j}_i}^2 \, {\cal H}_0^{ii}(r_1) \, = \, 
\int {\rm d}r_2 \, r_2^2 \, h_0(r_1,r_2) \, \rho(r_2) \, ,
\end{equation}
where
\begin{equation}
\label{density}
\rho(r) \, = \, \frac{1}{4\pi} \,\sum_i \, {\widehat{j}_i}^2 \,
|R_i(r)|^2 
\end{equation}
is the nuclear density.

For the three-point diagrams we have obtained the following
expressions: 
\begin{eqnarray}
D^{(\rm 1p1h)}_{3.1} & = & \sum_{i k L} \, \delta_{t_k,t_p} \,
\delta_{t_i,t_p} \, \delta_{j_i,j_p} \, \xi(l_p+l_k+L) \, \xi(l_p+l_i)
\, {\widehat{j}_k}^2 \, \threej{j_p}{j_k}{L}{\half}{-\half}{0}^2 \\ &&
\omega_{ih}^J  \,
\int {\rm d}r_1 \, r_1^2 \, R^*_i(r_1) \, 
F_J(qr_1) \,
R_h(r_1)
{\cal J}_L^{[pk];[ki]} \nonumber \, , \\[5mm] 
D^{(\rm 1p1h)}_{3.2} & = &
4\pi \, \sum_{i} \, \delta_{t_i,t_p} \, \delta_{j_i,j_p} \,
\xi(l_p+l_i) \, \omega_{ih}^J \,
\int {\rm d}r_1 \, r_1^2 \, R^*_i(r_1) \,
F_J(qr_1) \,
R_h(r_1) {\cal J}_0^{[pi];[\rho]} \, ,\\[5mm]
D^{(\rm 1p1h)}_{3.3} & = & \sum_{i k L} \, \delta_{t_i,t_h} \,
\delta_{t_k,t_h} \, \delta_{j_k,j_h} \, \xi(l_i+l_h+L) \, \xi(l_k+l_h)
\, {\widehat{j}_i}^2 \, \threej{j_i}{j_h}{L}{\half}{-\half}{0}^2 \\ &&
\omega_{pk}^J  \,
\int {\rm d}r_1 \, r_1^2 \, R^*_p(r_1) \, 
F_J(qr_1) \,
R_k(r_1) {\cal J}_L^{[ih];[ki]} \nonumber \, , \\[5mm] 
D^{(\rm 1p1h)}_{3.4} & = & 4\pi \, \sum_{i} \, \delta_{t_i,t_h} \, 
\delta_{j_i,j_h} \, \xi(l_i+l_h) \, 
\omega_{pi}^J \,  \int {\rm d}r_1 \, r_1^2 \, R^*_p(r_1) \,
F_J(qr_1) \,
R_i(r_1) {\cal J}_0^{[ih];[\rho]} \, ,\\[5mm]
D^{(\rm 1p1h)}_{3.5} & = & 4\pi \, (-1)^J \, \sum_{i k L} \,
\delta_{t_i,t_p} \, \delta_{t_k,t_h} \, (-1)^{j_i+j_k+L} \,
\frac{1}{{\widehat{L}}^2} \, \gamma_{pi}^L \, \gamma_{kh}^L \,
\sixj{j_p}{j_h}{J}{j_k}{j_i}{L}\\ && 
\omega_{ik}^J \,
\int {\rm d}r_1 \, r_1^2 \,
R^*_i(r_1) \, 
F_J(qr_1) \,
R_k(r_1) {\cal J}_L^{[pi];[kh]}
\nonumber \, , \\[5mm] 
D^{(\rm 1p1h)}_{3.6} & = & 4\pi \, \delta_{t_p,t_h}
\, \gamma_{ph}^J \, \frac{1}{{\widehat{J}}^4} \,\sum_{i k} \,
\delta_{t_i,t_k} \, \gamma_{ik}^J \,  \omega_{ik}^J \,
\int {\rm d}r_1
\, r_1^2 \, R^*_i(r_1) \, 
F_J(qr_1) \,
{\cal J}_J^{[ph];[ki]} \, .
\end{eqnarray}
\noindent
In these equations we have introduced the symbols:
\begin{equation}
\label{calJint}
{\cal J}_L^{[\alpha \beta];[\gamma \delta]} \, = \, 
\int {\rm d}r_1 \, r_1^2 \,
R^*_\alpha(r_1) \, R_\beta(r_1) \, {\cal H}_L^{[\gamma \delta]}(r_1) \, 
\end{equation}
and
\begin{equation}
\label{calJ0int}
{\cal J}_0^{[\alpha \beta];[\rho]} \, = \, 
\int {\rm d}r_1 \, r_1^2 \, R^*_\alpha(r_1) \, R_\beta(r_1) \, 
{\cal H}_0^{[\rho]}(r_1) \, .
\end{equation}

The equations presented above have been obtained for a generic
one-body transition operator of the form of Eq. (\ref{divop}). The
calculation of the electromagnetic response functions continues by
inserting in the above equations the explicit expressions 
for the charge and current operators, Eqs. 
(\ref{mjm})-(\ref{tmjm}). The final expressions are given in 
Appendix A.

In the case of two nucleon emission
the total angular momentum and parity of the nuclear final state
are given by the combination of the four angular momenta and the
parities of the particle and hole states involved in the excitation. 
We have chosen to couple the angular momenta of the two hole states
and of the two particle states first, and then, with another
recoupling, to obtain the total angular momentum of the 2p-2h excited
state: 
\begin{eqnarray}
\nonumber
|\Phi^{\Pi}_{JM}({\rm 2p2h})\rangle & = &
\sum_{M_p,M_h} \, \langle J_p M_p J_h -M_h | J M \rangle \,
\sum_{j_{\pon} m_{\pon} j_{\ptw} m_{\ptw}}\, 
\langle  j_{\pon} m_{\pon} j_{\ptw} m_{\ptw} |J_p M_p\rangle  \\ 
&& \hspace*{-1cm}\sum_{j_{\hon} m_{\hon} j_{\htw} m_{\htw}} \,
(-1)^{j_{\hon} + j_{\htw} - m_{\hon} - m_{\htw} }\,
\langle  j_{\hon} -m_{\hon} j_{\htw} -m_{\htw} |J_h M_h\rangle \, 
|\Phi^{\Pi}_{\rm 2p2h}\rangle \, .
\label{phi2p2h}
\end{eqnarray}
Inserting Eq. (\ref{phi2p2h}) into Eq. (\ref{xi2p2h})
and applying the Wigner-Eckart theorem 
we obtain the following expression:
\begin{equation}
\xi^{1,J}_{\rm 2p2h}(\bqu) \, = \, \langle \pon \ptw \| \,O_J (\bqu)
\,h_{12} \, \| \hon \htw \rangle \, +\, \sum^{A}_k \, \langle \pon
\ptw k \| \, O_J(\bqu) \, h_{23} \, | \hon \htw k\rangle \, ,
\label{xi2p2hj}
\end{equation}
where the coordinate labels have the same meaning as in
Eq. (\ref{xi1p1hj}). In the 2p-2h excitation
the energy conservation does not uniquely define the
energies of the emitted particles since
\begin{equation}
\omega\, = \, \epsilon_{\pon}\,+\,\epsilon_{\ptw}\,-\,
\epsilon_{\hon}\,-\,\epsilon_{\htw}\, ,
\end{equation}
for this reason in Eq. (\ref{phi2p2h}) an integration on the energy
of the particle $p_2$ is implied. 

It is evident from Fig. 2 that the various diagrams
can be related one to the other by changing the labels of the hole and
of the particle states involved. 
We use the following symmetry relations to
calculate the contribution of the various diagrams to the
response:
\begin{eqnarray}
\nonumber
D^{(\rm 2p2h)}_{2.2}[\pon \hon \ptw \htw] & = &
(-1)^{j_{\hon}+j_{\htw}+J_h} 
D^{(\rm 2p2h)}_{2.1}[\pon \htw \ptw \hon] \, , \\
D^{(\rm 2p2h)}_{2.3}[\pon \hon \ptw \htw] & = & (-1)^{j_{\pon}+j_{\ptw}+J_p} \,
D^{(\rm 2p2h)}_{2.1}[\ptw \hon \pon \htw] \, , \\
\nonumber
D^{(\rm 2p2h)}_{2.4}[\pon \hon \ptw \htw] & = & (-1)^{j_{\pon}+j_{\ptw}+J_p} \,
(-1)^{j_{\hon}+j_{\htw}+J_h} \, D^{(\rm 2p2h)}_{2.1}[\ptw \htw \pon \hon] .
\end{eqnarray}
The explicit expression of $D^{(\rm 2p2h)}_{2.1}$
matrix element for a generic one-body operator is given by:
\begin{eqnarray}
D^{(\rm 2p2h)}_{2.1}[\pon \hon \ptw \htw] & = & 
4\pi \, \delta_{t_{\ptw},t_{\htw}} \, \widehat{J}_p \,
\widehat{J}_h \, \sum_{L K} \, (-1)^L \, \frac{\widehat{K}}{{\widehat{L}}^2} 
\, \gamma_{{\ptw} {\htw}}^L \,
\ninej{j_{\pon}}{j_{\hon}}{K}{j_{\ptw}}{j_{\htw}}{L}{J_p}{J_h}{J} \\
&&\nonumber 
\int {\rm d}r_1 \, r_1^2 \, R^*_{\pon}(r_1) \, \omega_{{\pon} {\hon}}^{LJ;K} \,
R_{\hon}(r_1) \, {\cal H}_L^{[{\ptw} {\htw}]}(r_1) \, .
\end{eqnarray}

The symmetry relations we used to evaluate the 
the three-point diagrams are:
\begin{eqnarray}
\nonumber
D^{(\rm 2p2h)}_{3.2}[\pon \hon \ptw \htw] & = & 
(-1)^{j_{\hon}+j_{\htw}+J_h} \,
D^{(\rm 2p2h)}_{3.1}[\pon \htw \ptw \hon] \, , \\
\nonumber
D^{(\rm 2p2h)}_{3.3}[\pon \hon \ptw \htw] & = & 
(-1)^{j_{\pon}+j_{\ptw}+J_p} \,
D^{(\rm 2p2h)}_{3.1}[\ptw \hon \pon \htw] \, , \\
\nonumber
D^{(\rm 2p2h)}_{3.4}[\pon \hon \ptw \htw] & = & 
(-1)^{j_{\pon}+j_{\ptw}+J_p} \,
(-1)^{j_{\hon}+j_{\htw}+J_h} \, 
D^{(\rm 2p2h)}_{3.1}[\ptw \htw \pon \hon] \, ,\\
D^{(\rm 2p2h)}_{3.6}[\pon \hon \ptw \htw] & = & 
(-1)^{j_{\hon}+j_{\htw}+J_h} \,
D^{(\rm 2p2h)}_{3.5}[\pon \htw \ptw \hon] \, , \\
\nonumber
D^{(\rm 2p2h)}_{3.7}[\pon \hon \ptw \htw] & = & 
(-1)^{j_{\pon}+j_{\ptw}+J_p} \,
D^{(\rm 2p2h)}_{3.5}[\ptw \hon \pon \htw] \, , \\
\nonumber
D^{(\rm 2p2h)}_{3.8}[\pon \hon \ptw \htw] & = & 
(-1)^{j_{\pon}+j_{\ptw}+J_p} \,
(-1)^{j_{\hon}+j_{\htw}+J_h} \, 
D^{(\rm 2p2h)}_{3.5}[\ptw \htw \pon \hon]
\, , \\
\nonumber
D^{(\rm 2p2h)}_{3.10}[\pon \hon \ptw \htw] & = & 
(-1)^{j_{\pon}+j_{\ptw}+J_p} \,
(-1)^{j_{\hon}+j_{\htw}+J_h} \, 
D^{(\rm 2p2h)}_{3.9}[\ptw \htw \pon \hon]
\, , \\
\nonumber
D^{(\rm 2p2h)}_{3.12}[\pon \hon \ptw \htw] & = & 
(-1)^{j_{\pon}+j_{\ptw}+J_p} \,
(-1)^{j_{\hon}+j_{\htw}+J_h} \, 
D^{(\rm 2p2h)}_{3.11}[\ptw \htw \pon \hon]
\, .
\end{eqnarray}
The corresponding matrix elements for a generic one-body operator are:
\begin{eqnarray}
D^{(\rm 2p2h)}_{3.1}[\pon \hon \ptw \htw] & = & 
4\pi \, \delta_{t_{\ptw},t_{\htw}} \,
(-1)^{j_{\pon}+j_{\hon}+J+1} \, \widehat{J}_p \, \widehat{J}_h 
\, \sum_{i L} \,  \delta_{t_i,t_{\hon}} \, \frac{1}{{\widehat{L}}^2} \,
\gamma_{{\ptw} {\htw}}^L \, \gamma_{i {\hon}}^L \\
&& \hspace*{-1.5cm}\sixj{j_{\ptw}}{J_h}{j_i}{j_{\hon}}{L}{j_{\htw}} \,
\sixj{J_p}{J_h}{J}{j_i}{j_{\pon}}{j_{\ptw}} \, 
\int {\rm d}r_1 \, r_1^2 \, R^*_{\pon}(r_1) \, \omega_{{\pon} i}^J \,
R_i(r_1) \, {\cal J}_L^{[{\ptw} {\htw}];[i {\hon}]} \nonumber \, ,
\\[5mm]
D^{(\rm 2p2h)}_{3.5}[\pon \hon \ptw \htw]  & = & 
4\pi \, \delta_{t_{\pon},t_{\hon}} \,
(-1)^{j_{\ptw}+j_{\htw}+J+1} \, \widehat{J}_p \, \widehat{J}_h 
\, \sum_{i L} \,  \delta_{t_i,t_{\ptw}} \, \frac{1}{{\widehat{L}}^2} \,
\gamma_{{\pon} {\hon}}^L \, \gamma_{{\ptw} i}^L \\
&&  \hspace*{-1.5cm} \sixj{j_i}{j_{\hon}}{J_p}{j_{\pon}}{j_{\ptw}}{L} \,
\sixj{J_p}{J_h}{J}{j_{\htw}}{j_i}{j_{\hon}} \, 
\int {\rm d}r_1 \, r_1^2 \, R^*_i(r_1) \, \omega_{i {\htw}}^J \,
R_{\htw}(r_1) \, {\cal J}_L^{[{\pon} {\hon}];[{\ptw} i]} \nonumber \, ,
\\[5mm]
D^{(\rm 2p2h)}_{3.9}[\pon \hon \ptw \htw]  & = & 
4\pi \, \delta_{t_{\ptw},t_{\hon}} \,
\delta_{j_{\ptw},j_{\hon}} \, (-1)^{j_{\pon}+j_{\ptw}+j_{\htw}+J-\half} 
\, \frac{\widehat{J}_p \, \widehat{J}_h}{{\widehat{j}_{\ptw}}^2} \,
\sixj{J_p}{J_h}{J}{j_{\htw}}{j_{\pon}}{j_{\ptw}} \\
&& \hspace*{-1.5cm} \int {\rm d}r_1 \, r_1^2 \, R^*_{\pon}(r_1) \, 
\omega_{{\pon} {\htw}}^J \, R_{\htw}(r_1) \,
\sum_{iL} \, \delta_{t_i,t_{\ptw}} \, (-1)^{j_i+\half} \, \frac{1}
{{\widehat{L}}^2} \,\gamma_{{\ptw} i}^L \,\gamma_{i {\hon}}^L 
\, {\cal J}_L^{[{\ptw} i];[i {\hon} ]} \nonumber \, ,
\\[5mm]
D^{(\rm 2p2h)}_{3.11}[\pon \hon \ptw \htw]  & = & 
4\pi \, \delta_{t_{\ptw},t_{\hon}} \,
\delta_{j_{\ptw},j_{\hon}} \, (-1)^{j_{\pon}+j_{\htw}+J+1}
\, \widehat{J}_p \, \widehat{J}_h \, \xi(l_{\ptw}+l_{\hon}) \\
&&  \hspace*{-1.5cm} \sixj{J_p}{J_h}{J}{j_{\htw}}{j_{\pon}}{j_{\ptw}} \, 
\int {\rm d}r_1 \, r_1^2 \, R^*_{\pon}(r_1) \, 
\omega_{{\pon} {\htw}}^J \, R_{\htw}(r_1)
\, {\cal J}_0^{[{\ptw} {\hon}];[\rho]} \nonumber \, .
\end{eqnarray}

Also in this case the contribution of the electromagnetic charge and
current operator to the two nucleon emission responses has to be
calculated by inserting  Eqs. (\ref{mjm})-(\ref{tmjm}) in the
above equations. The explicit expressions we have obtained are
given in Appendix B.

\section{Specific applications}
\label{sect:sa}

The model described above has been applied to evaluate electromagnetic
responses and cross sections in the quasi-elastic region for the
doubly magic nuclei $^{16}$O and $^{40}$Ca. 

The two-body correlation function and the
single particle basis are 
the two inputs necessary to build the many-body wavefunctions
used in the calculation of the responses. 
In our calculations the set of single particle wave functions 
have been generated by a spherical Woods-Saxon potential of the form: 
\begin{equation}
V(r) \,= \, {V_0} F(r) \,
+ \, \left (\frac {\hbar c}{m_{\pi}} \right)^2\, 
\frac{V_{LS}}{r} \, \frac{\rm d}
{{\rm    d}r} F(r) 
\,\, {\bf l} \cdot \bsigma \, + \, V_C(r)  \, ,
\end{equation}
where $m_{\pi}$ is the pion mass, $V_C(r)$ the Coulomb potential
produced by a homogeneously charged sphere of radius $R$, and
\begin{equation}
 F(r)\, = \, \frac{1} {1+e^{(r-R)/a} } \, .
\end{equation}
The same parameterization of the potential has been used 
for both bound and continuum waves.

As we have already anticipated,
the other input of our approach, the two-body correlation function,  
is a purely scalar function.
Single particle waves and correlation functions are not
independent quantities. 
In the theoretical framework of the CBF theory they  
are fixed by the minimization of the ground state energy functional. 
In our calculations we use wave functions and correlations
taken from CBF-FHNC calculations of $^{16}$O and $^{40}$Ca
ground states done with different hamiltonians.

A first set of many-body
wavefunctions has obtained in Ref. \cite{ari96}
with the semi-realistic S3 interaction of Afnan and Tang.
Specifically, we used the Woods-Saxon parameters of Tab. 5 of that
reference together with the correlations labelled Euler in Tab. 6.
We should recall that in that calculation
the Woods-Saxon parameterizations was taken from the literature and
the energy minimum was obtained by changing only the correlation
function. 

The second set of wave functions has been taken from 
Ref.\cite{fab00} where a hamiltonian containing
a V8' Argonne potential, which is a truncated
version of the V18 interaction, together with the Urbana IX
three-body interaction has been used.
Our calculations have been done with
the wavefunctions of Tab. V of that reference. Also in this
case the Woods-Saxon parameters have been kept fixed and the energy
minimum was found by changing only the correlation function. 
We should remark however 
that the energy minima found in these restricted
calculations are only 7\% larger than the those found by the
full minimization. 
In our calculations we used only the scalar terms
of the correlation functions of Ref. \cite{fab00} which consider
the channels of Eq. (\ref{oper}) up to $p=6$.

\begin{table}
\caption{Parameters of the Woods-Saxon potentials used to generate the
  set of single particle wave functions.}
\label{tab:pot}
\begin{center}
\begin{tabular}{clrrrr}
\hline
 & & \multicolumn{2}{c}{$^{16}$O} & \multicolumn{2}{c}{$^{40}$Ca} \\
\cline{3-6}
 &  &  V8    & S3   &  V8    & S3     \\ 
\hline
 $\pi$ & $V_0$  [MeV]   & -53.0 & -52.5  & -50.0 & -57.5 \\
       & $V_{ls}$ [MeV] & 0.0   &  -7.0  &   0.0 & -11.11  \\
       & $R$   [fm]     &  3.45 &  3.2   &   4.6 &  4.1  \\ 
       & $a$   [fm]     &  0.7  &  0.53  &   0.5 &  0.53 \\
\hline
 $\nu$ & $V_0$  [MeV]   & -53.0 & -52.5  & -50.0 & -55.0 \\
       & $V_{ls}$ [MeV] & 0.0   &  -6.54 &   0.0 & -8.5  \\
       & $R$   [fm]     &  3.45 &  3.2   &   4.6 &  4.1  \\ 
       & $a$   [fm]     &  0.7  &  0.53  &   0.5 &  0.53 \\
\hline
\end{tabular}
\end{center}
\end{table}

\begin{table}
\caption{Single particle energies, in MeV, of the $^{16}$O
nucleus obtained with the two chosen potentials. 
We have indicated with $\pi$ and $\nu$ proton and neutron states 
respectively.}
\label{tab:speo16}
\begin{center}
\begin{tabular}{clrrr}
\hline
 &  &
\multicolumn{3}{c}{$^{16}$O} \\
\cline{3-5}
 &  &  V8   & S3     & exp \\ 
\hline
 $\pi$          & 1s1/2 &  -30.7 & -30.0 &        \\
                & 1p3/2 &  -17.6 & -16.9 & -18.4 \\
                & 1p1/2 &  -17.6 & -12.7 & -12.1 \\
\hline
 $\nu$          & 1s1/2 &  -34.5 & -34.0 &       \\
                & 1p3/2 &  -21.0 & -20.5 & -21.8 \\
                & 1p1/2 &  -21.0 & -16.6 & -15.7 \\
\hline
\end{tabular}
\end{center}
\end{table}

\begin{table}
\caption{Single particle energies, in MeV, of the $^{40}$Ca nucleus
obtained with the two chosen potentials.  We have indicated with $\pi$
and $\nu$ proton and neutron states respectively.}
\label{tab:speca40}
\begin{center}
\begin{tabular}{clrrr}
\hline
 & & \multicolumn{3}{c}{$^{40}$Ca} \\
\cline{3-5}
  & &  V8    & S3  & exp \\ 
\hline
 $\pi$          & 1s1/2 & -32.5 & -36.0 &     \\
                & 1p3/2 & -24.0 & -26.5 &     \\
                & 1p1/2 & -24.0 & -23.0 &     \\
                & 1d5/2 & -14.0 & -16.2 & -14.7\\
                & 1d3/2 & -14.0 &  -8.7 &  -8.3\\
                & 2s1/2 & -11.0 & -10.3 & -10.3 \\
\hline
 $\nu$          & 1s1/2 & -40.4 & -42.6 & \\
                & 1p3/2 & -31.5 & -32.4 & \\
                & 1p1/2 & -31.5 & -29.8 & \\
                & 1d5/2 & -21.1 & -21.4 & -18.2\\
                & 1d3/2 & -21.1 & -15.7 & -15.6\\
                & 2s1/2 & -18.2 & -16.4 & -21.6\\
\hline
\end{tabular}
\end{center}
\end{table}

\begin{figure}
\begin{center}
\hspace*{-2.0cm}
\leavevmode
\epsfysize = 400pt
\epsfbox[70 200 500 650]{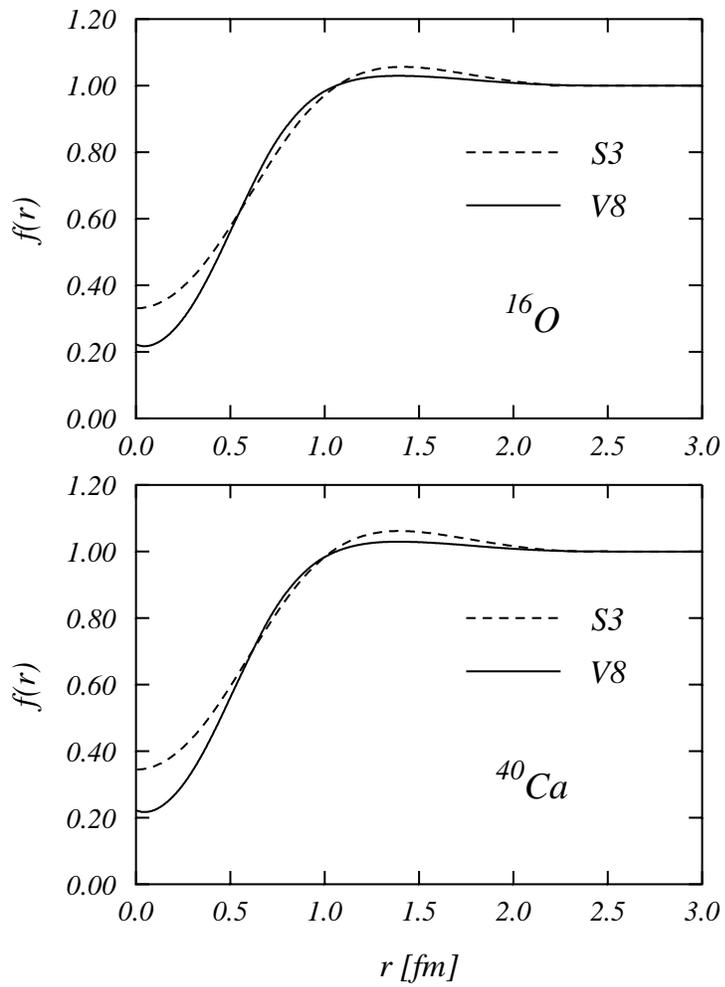}
\end{center}
\vspace{-.5cm}
\caption{\small Correlation functions used in the calculations.  The
label S3 refer to the FHNC calculations of Ref. \protect\cite{ari96}
while V8 to those of Ref. \protect\cite{fab00}.}
\label{fig:corr}
\end{figure}

Henceforth we shall label S3 and V8 all the quantities related to the
wavefunctions of Ref.\cite{ari96} and \cite{fab00} respectively.
We show in Fig. 3 the correlation functions $f(r)$
and we give in Tab. \ref{tab:pot} the values of the parameters of the
Woods-Saxon well.
To complete the information on the input, we present in
Tab. \ref{tab:speo16}  and \ref{tab:speca40} 
the values of the single particle energies 
and we compare in Fig. 4 the MF and the
correlated charge density distributions
with the empirical densities of Ref. \cite{dej87}.
The charge densities have been obtained by folding the pointlike
proton densities with the electromagnetic nucleon form factor of 
Ref. \cite{hoe76}. 
The same nucleon form factor has been used in the calculation of the
responses.

\begin{figure}
\begin{center}
\hspace*{-2.0 cm}
\leavevmode
\epsfysize = 300pt
\epsfbox[70 200 500 650]{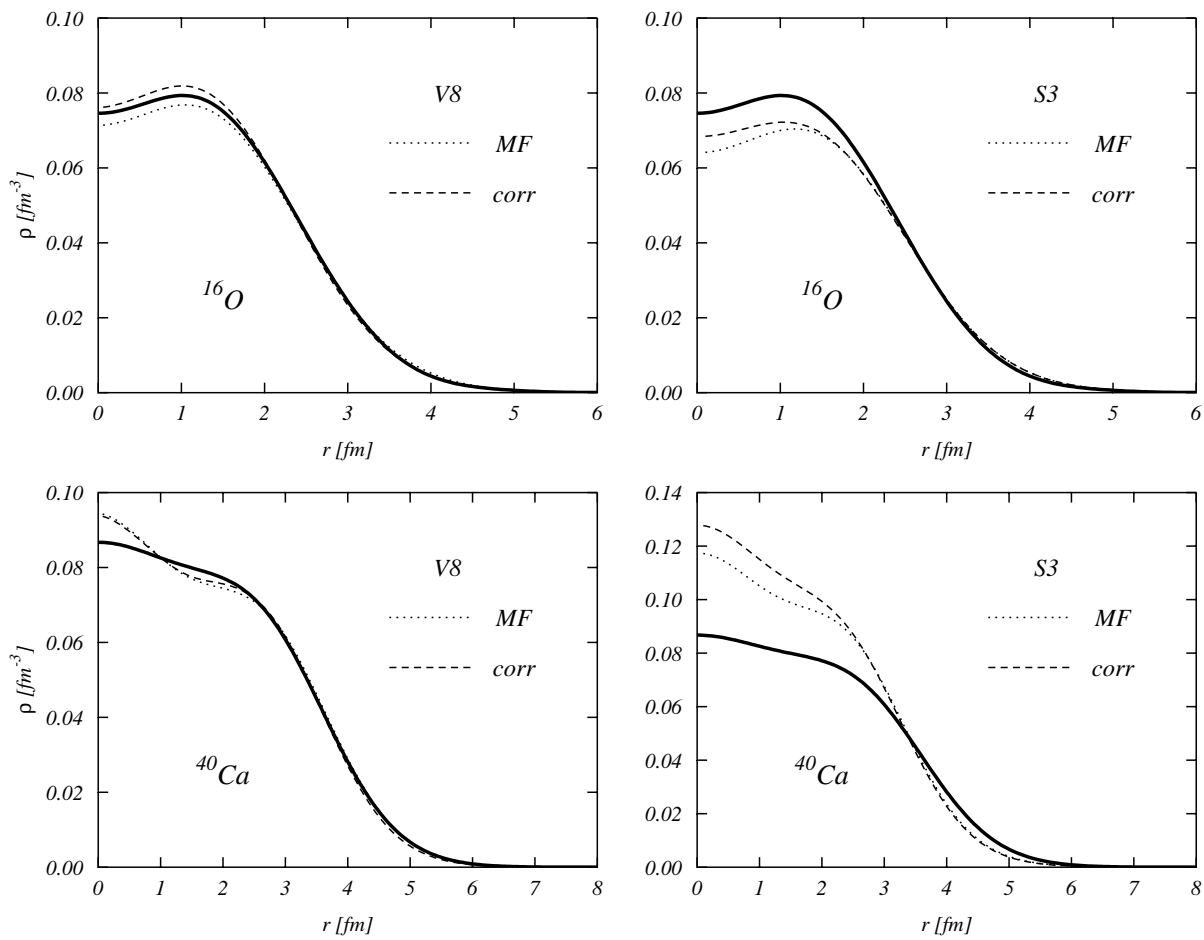}
\end{center}
\vspace{3.0cm}
\caption{ \small MF (dotted lines) and correlated (dashed lines)
charge densities distributions compared with the empirical densities
(full lines) taken from the compilation of
Ref. \protect\cite{dej87}. }
\label{fig:den4}
\end{figure}

Since the convection current contribution is already small at the MF
level, as we show in Fig. 5, we did not
consider the correlated terms produced by this current.
In any case the convection current is included in the mean-field (MF)
term. 

\begin{figure}
\begin{center}
\hspace*{-2.0cm}
\leavevmode
\epsfysize = 400pt
\epsfbox[70 200 500 650]{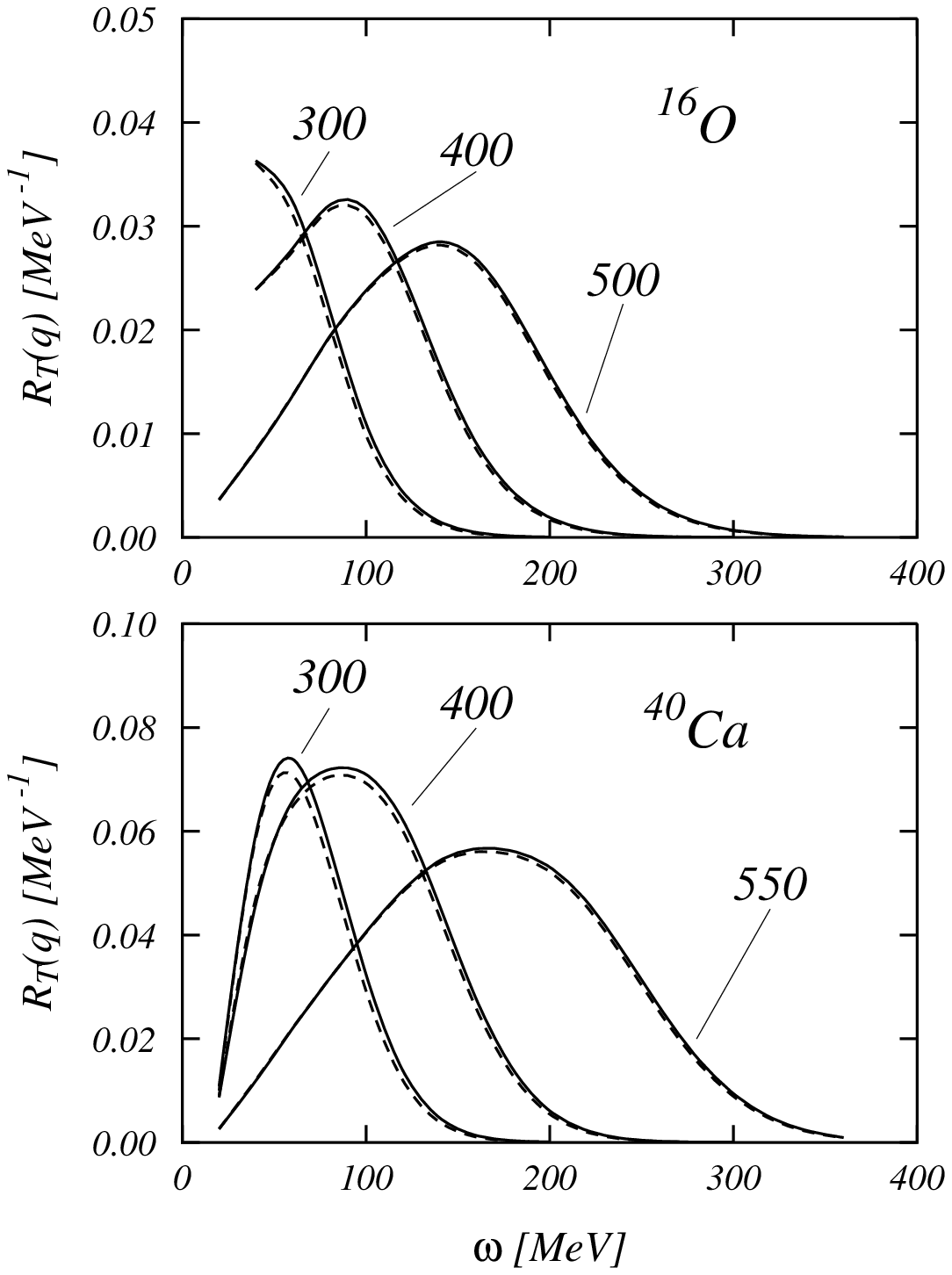}
\end{center}
\caption{ \small MF transverse responses calculated with the S3
Woods-Saxon potentials considering magnetization current only (dashed
lines) and magnetization plus convection current (full lines). The
numbers in the figure indicates the values of the momentum transfer in
MeV/$c$ units. }
\label{fig:conv}
\end{figure}

\subsection{One-nucleon emission}
In Figs. 6, 7, 8, and 9 we present the electromagnetic 1p-1h responses
calculated for the two nuclei considered for three different values of
the the momentum transfer.

\begin{figure}
\begin{center}
\hspace*{-2.0 cm}
\leavevmode
\epsfysize = 300pt
\epsfbox[70 200 500 650]{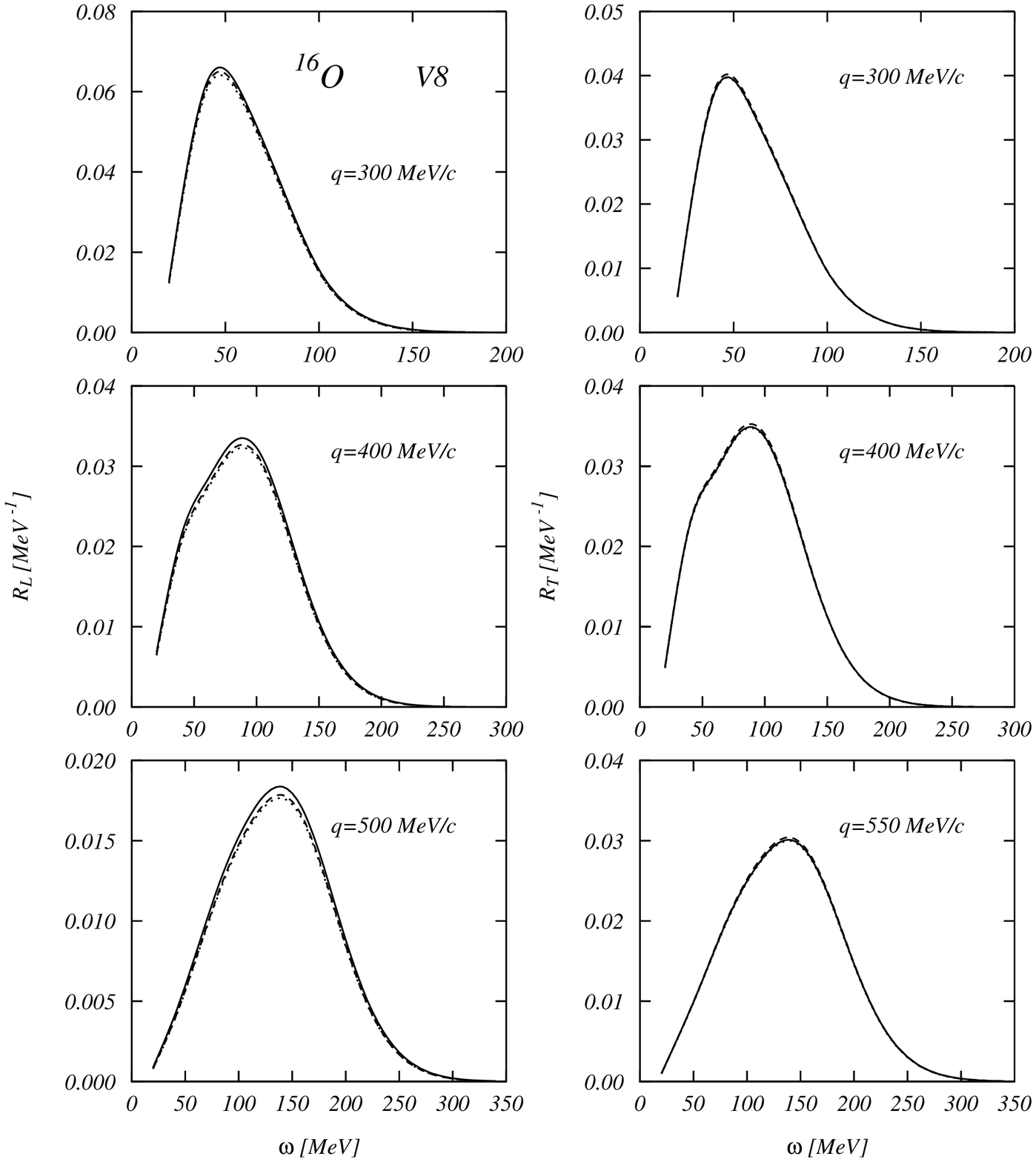}
\end{center}
\vspace{2.0cm}
\caption{\small Longitudinal and transverse 1p-1h responses in
$^{16}$O for three different values of the momentum transfer.  The
calculations have been done with the V8 wave functions and
correlations.  The full lines show the MF responses, the dotted lines
have been obtained with the inclusion of the two-point diagrams of
Fig. \protect\ref{fig:diag1p1h}, while the dashed lines have been
obtained considering all the diagrams of that figure. }
\label{fig:rv8o16}
\end{figure}

\begin{figure}
\begin{center}
\hspace*{-2.0 cm}
\leavevmode
\epsfysize = 300pt
\epsfbox[70 200 500 650]{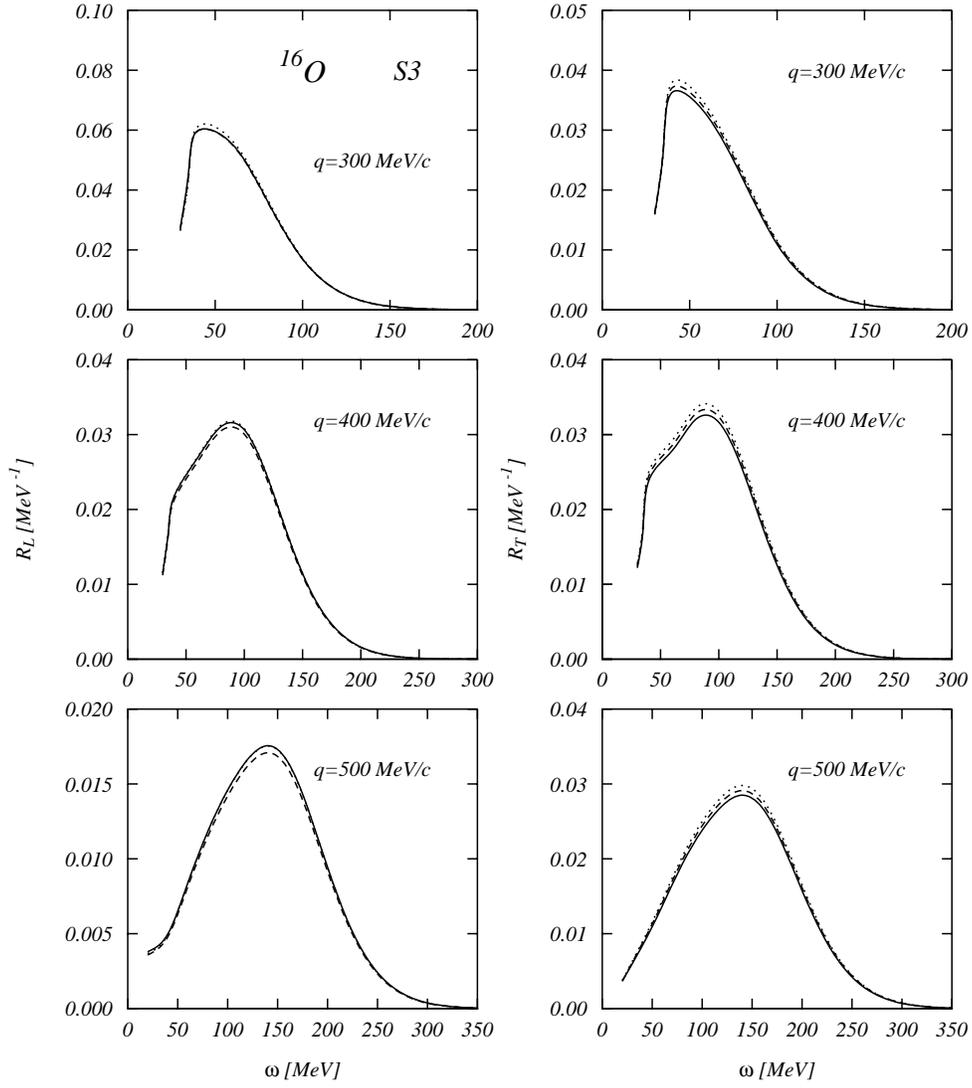}
\end{center}
\vspace{2.0cm}
\caption{\small The same as in Fig. \protect\ref{fig:rv8o16} with the
S3 input. }
\label{fig:rs3o16}
\end{figure}

\begin{figure}
\begin{center}
\hspace*{-2.0 cm}
\leavevmode
\epsfysize = 300pt
\epsfbox[70 200 500 650]{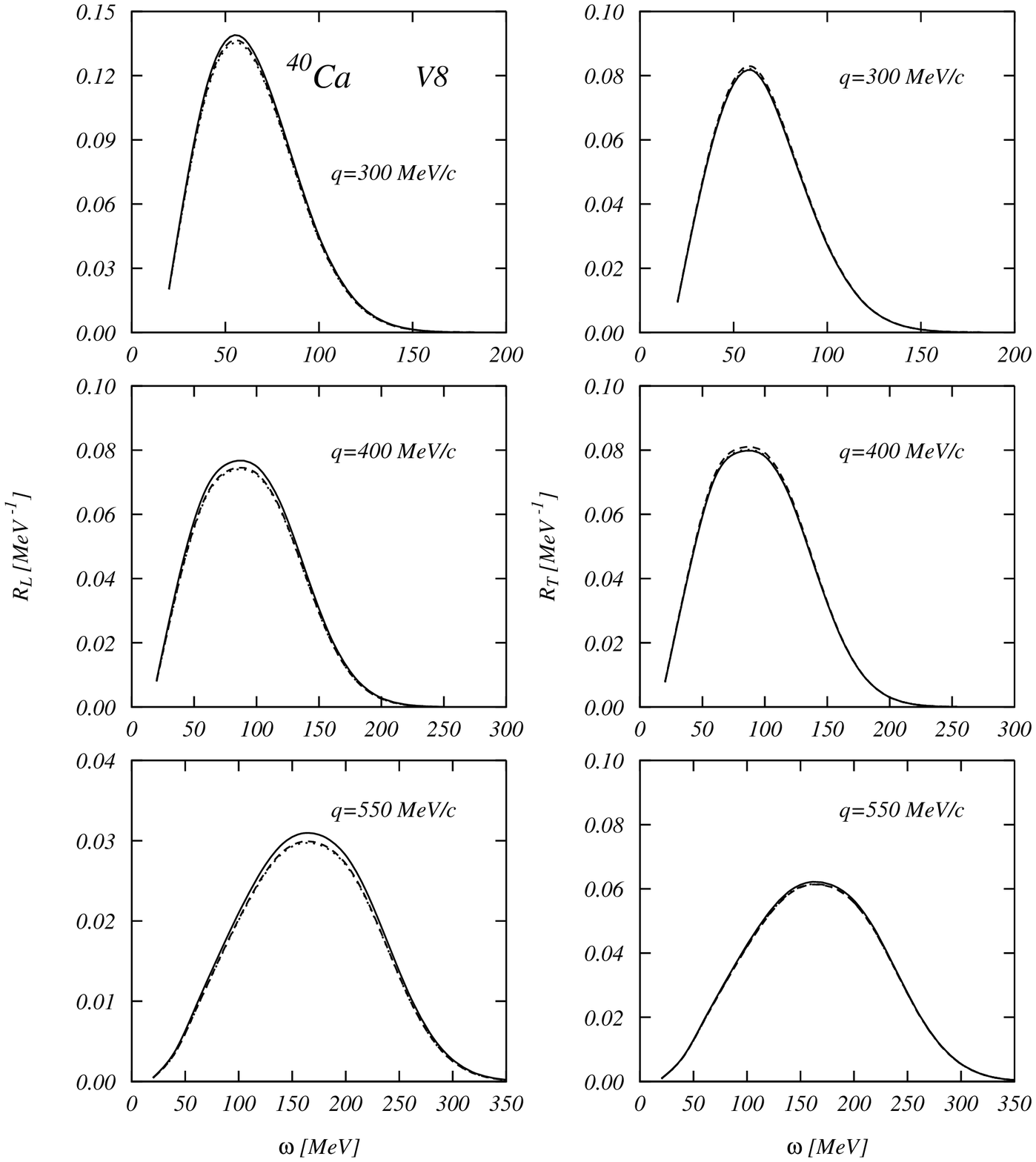}
\end{center}
\vspace{2.0cm}
\caption{\small The same as in Fig.  \protect\ref{fig:rv8o16} for
$^{40}$Ca.  }
\label{fig:rv8ca40}
\end{figure}

\begin{figure}
\begin{center}
\hspace*{-2.0 cm}
\leavevmode
\epsfysize = 300pt
\epsfbox[70 200 500 650]{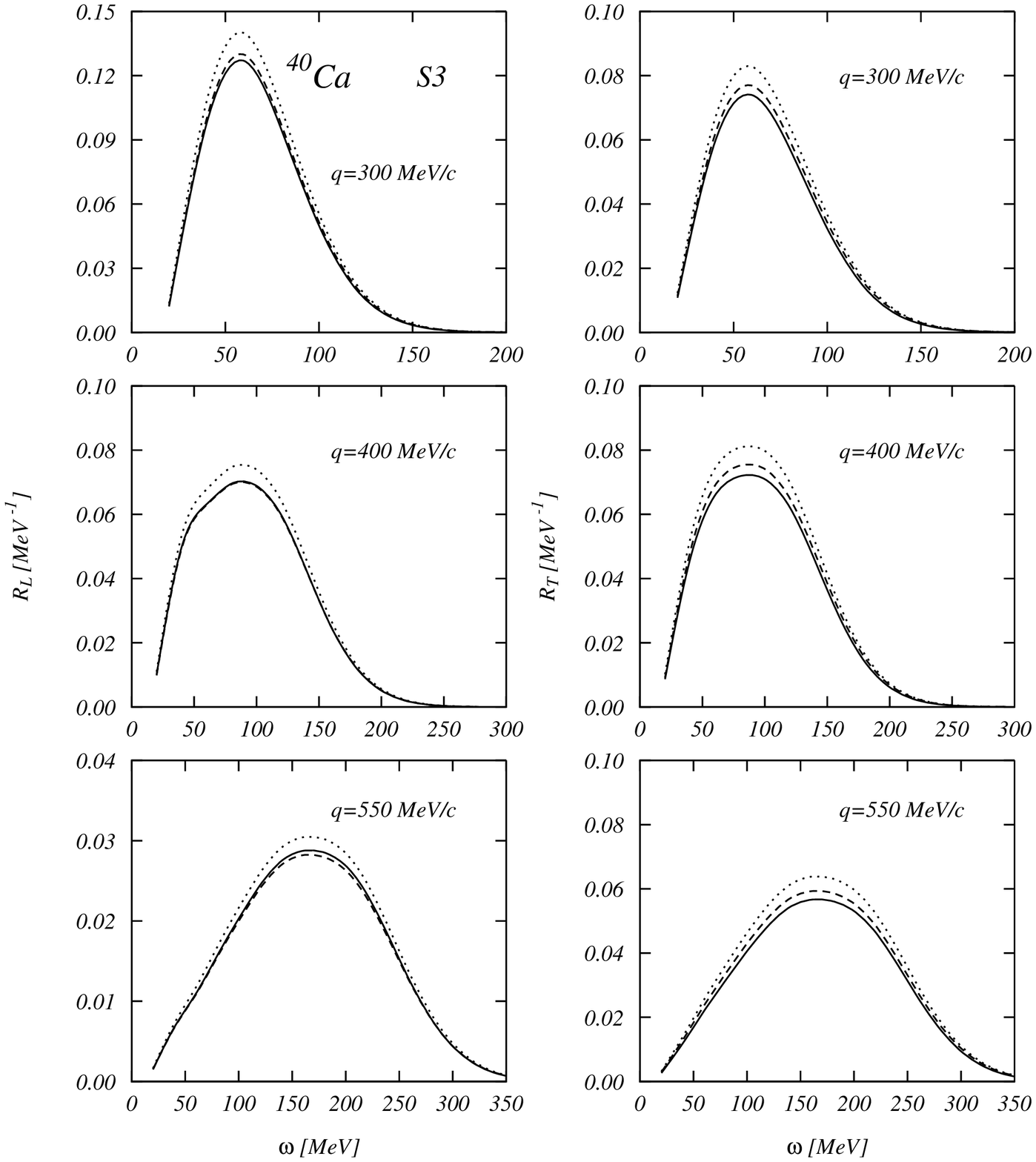}
\end{center}
\vspace{2.0cm}
\caption{\small The same as in Fig.  \protect\ref{fig:rs3o16} for
$^{40}$Ca. }
\label{fig:rs3ca40}
\end{figure}

In these figures the full lines indicate the MF results. 
The dotted lines have been obtained by
adding to the MF responses the contribution of
two-point diagrams of
Fig. 1, while the dashed lines show 
the results of the calculations when all the diagrams
have been considered.

It is evident that the short-range correlations produce small effects
on the inclusive responses. In the peak positions we found maximum
relative variations with respect to the MF responses of 1.7\% in the
longitudinal and of 2.2\% in the transverse responses.  These values
are within the range of the uncertainty produced by the different
choices of nucleon form factors \cite{ama93}.
In spite of this we made a detailed investigation of the correlation
effects because they can be relevant in the study of exclusive
experiments.  

\begin{figure}
\begin{center}
\hspace*{-2.0 cm}
\leavevmode
\epsfysize = 300pt
\epsfbox[70 200 500 650]{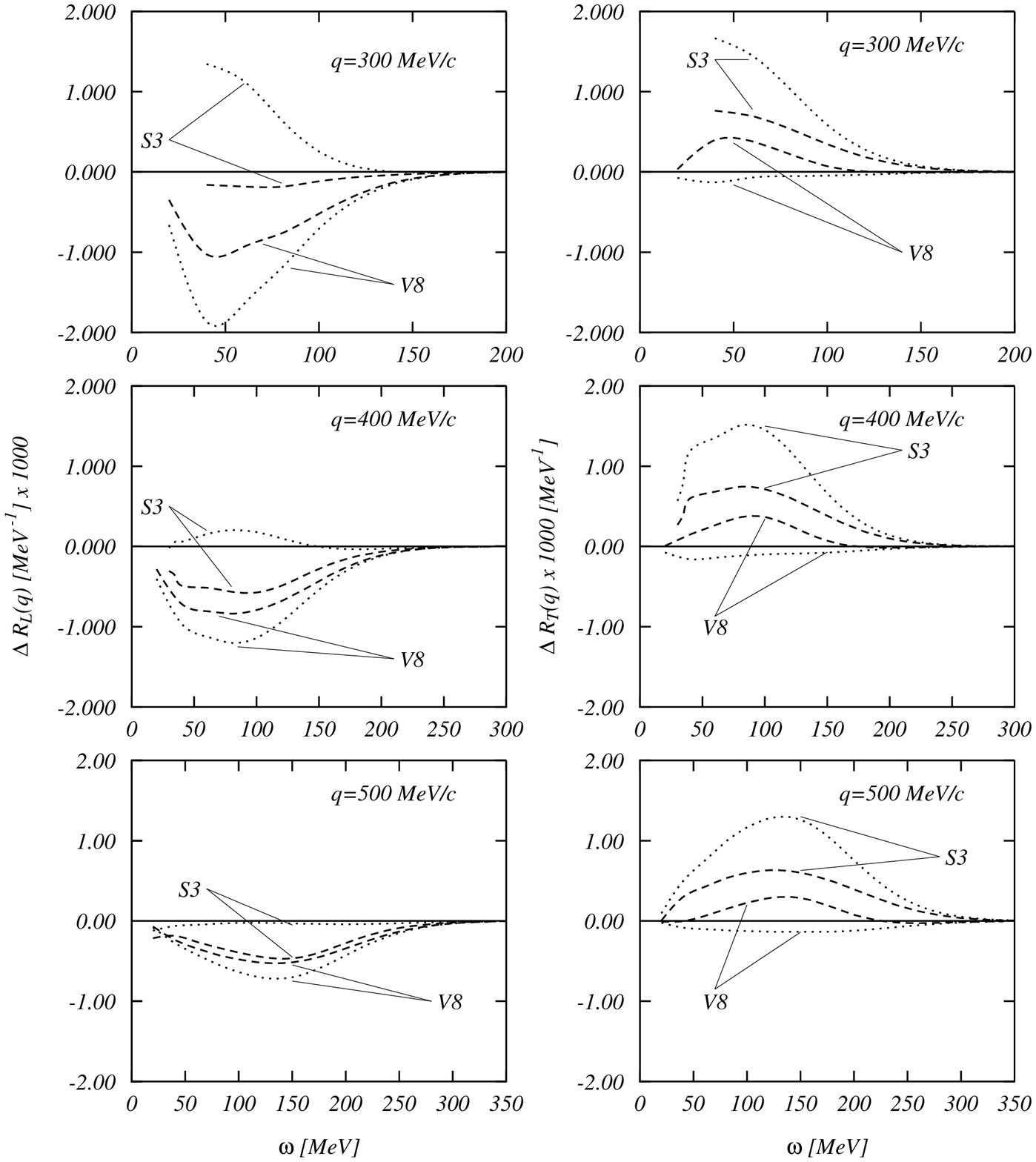}
\end{center}
\vspace{2.0cm}
\caption{\small Differences between correlated and MF responses as
defined in Eq. (\protect\ref{eq:deltaLT}) for $^{16}$O.}
\label{fig:diffo16}
\end{figure}

\begin{figure}
\begin{center}
\hspace*{-2.0cm}
\leavevmode
\epsfysize = 300pt
\epsfbox[70 200 500 650]{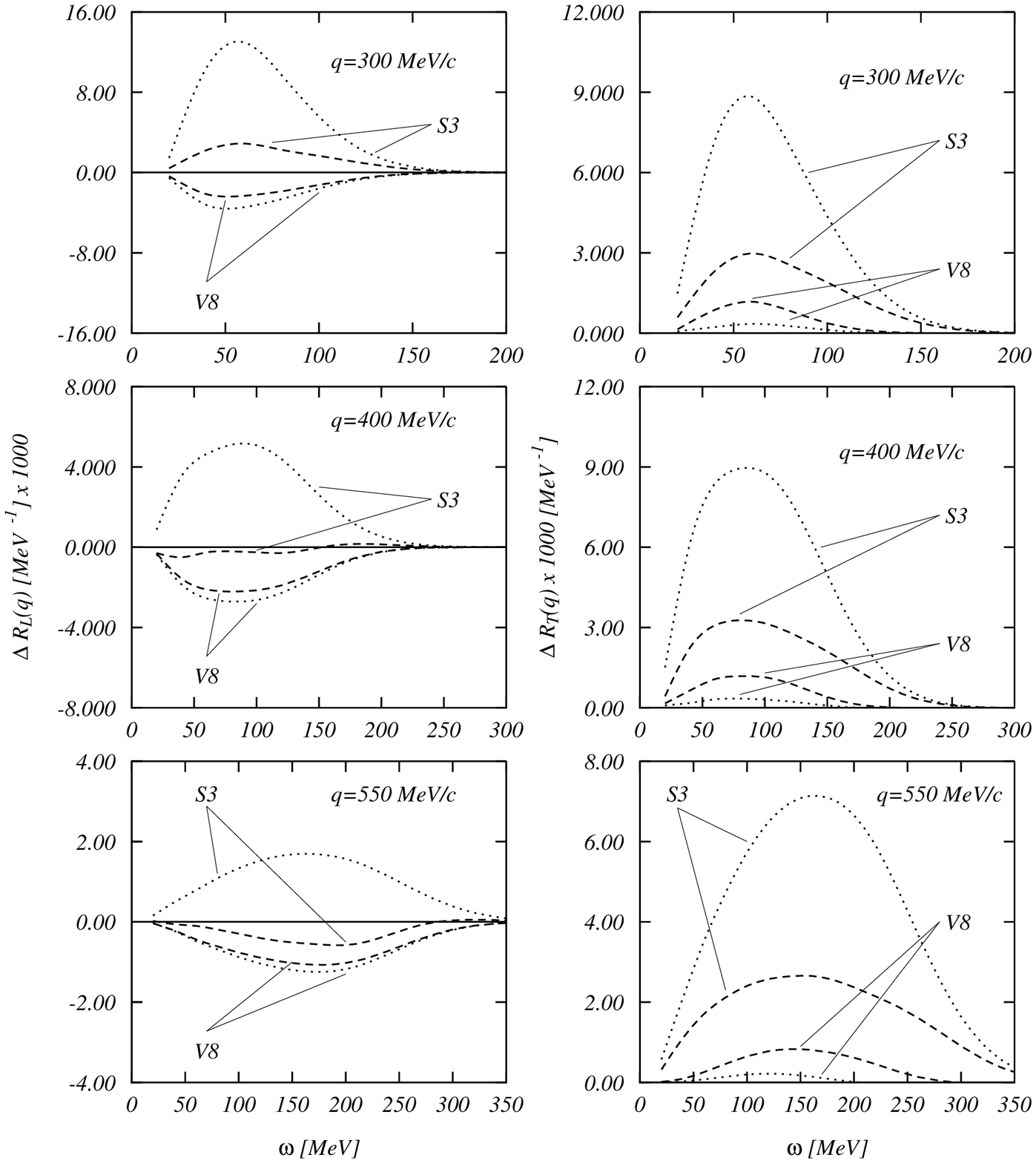}
\end{center}
\vspace{2.0cm}
\caption{\small The same as Fig. \protect\ref{fig:diffo16} for
$^{40}$Ca. }
\label{fig:diffca40}
\end{figure}

In order to have a better view of the correlation effects we show in 
Figs. 10 and 11 
the difference between the correlated and MF responses:
\begin{equation}
\Delta R_{L,T} (q,\omega)\, = 
\, R_{L,T}(q,\omega) - R_{L,T}^{\rm MF}(q,\omega) \, .
\label{eq:deltaLT}
\end{equation}
In these figures we have used 
the convention of indicating with the dotted lines
the results obtained with the two-point diagrams only, and with the
dashed line those obtained with all the diagrams.

The inclusion of the three-point diagrams reduces the effects produced 
by the two-point diagrams alone.
This fact is more evident in the
longitudinal responses and it is consistent with the findings of
Refs. \cite{co95,ari97} where ground state charge and momentum
distributions have been studied.  As discussed in
sect. \ref{sect:model}, the proper normalization of the many-body wave
functions is obtained because in the limit for q $\rightarrow$ 0 the
three-point diagrams cancel exactly the contribution of the two-point
diagrams. This cancellation is present also for large values of the
momentum transfer, but it is not exact any more. It is interesting to
notice that the final correlated longitudinal responses are always
smaller than the MF responses with the only exception of the S3 result
in $^{40}$Ca at 300 MeV/$c$.

The behavior of the correlations on the transverse responses is
rather different: the correlated responses are always larger than the
MF ones. The reason of this behavior has to be ascribed to the
diagram (2.3) of Fig. 1 which does not contributes
in the transverse response as indicated in Appendix A.  In this
diagram the exchanged virtual photon strikes a nucleon which remains
bound but through the correlation it excites a p-h pair producing the
nucleon emission.  Since in our calculations the correlation is a
scalar function, only longitudinally polarized virtual photons can
induce this type of excitation.

\begin{figure}
\begin{center}
\hspace*{-2.0cm}
\leavevmode
\epsfysize = 350pt
\epsfbox[70 200 500 650]{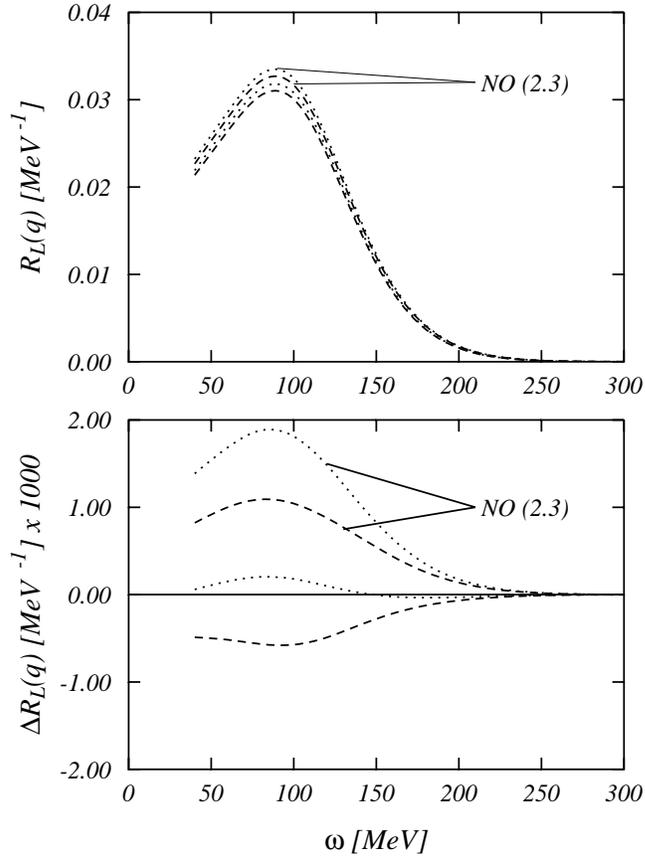}
\end{center}
\caption{\small Upper panel: $^{16}$O longitudinal responses
calculated with the S3 input.  The curves indicated with {\bf NO
(2.3)} have been obtained without the (2.3) diagram of
Fig. \protect\ref{fig:diag1p1h}. The other curves correspond to those
shown in Fig.  \protect\ref{fig:rs3o16}.  Lower panel: difference
between the correlated and MF responses.  Also in this figure the
dotted lines indicate the results obtained by using only the two-point
diagrams, while the dashed lines show the results of the full
calculation. }
\label{fig:no23}
\end{figure}

To study the effect of this diagram we set it
artificially to zero also in the longitudinal channel. The comparison
with the correctly calculated longitudinal response is shown in
Fig. 12 for the $^{16}$O response at $q=400$~MeV/$c$
calculated with the S3 wavefunctions.  In the upper panel we show the
responses, and in the lower one the $\Delta R_l$ of 
Eq. (\ref{eq:deltaLT}). The (2.3) diagram lowers the response and
produces the discussed difference between longitudinal and transverse
responses. The curves without the contribution of the (2.3) diagram in
the lower panel are very similar to the analogous ones of
Fig. 10 for the transverse responses.

Another interesting feature observed in Figs. 10 and 11 is the
different behavior shown by the S3 and V8 correlations. The two-point
contributions of the S3 correlation are always positive, and the S3
total contributions are always larger than those of the V8.

The different features characterizing the two correlation functions
are shown in Fig. 3.
A first difference concerns the behavior
at $r=0$ where the  S3 correlations assume larger values than the V8
ones. The second difference is related to the behavior at $r$ values
around 1-1.5~fm, where the S3 correlation functions show a sizable
overshooting with respect to the asymptotic value of 1. 
In order to study the effects of these two features  on the response
functions, we performed calculations using correlation functions of
the form:
\begin{equation} 
f(r)=1 - A \exp[-B r^2] + C \exp[ - D (r-r_0)^2 ] \, .
\label{gcorr}
\end{equation}
This expression allowed us to disentangle and magnify the
characteristics of the correlation function under investigation by
changing the values of the parameters. The values used in our
calculations are given in Table \ref{tab:gcorr}.  The correlations g1,
g2 and g3, shown in the upper panel of Fig. 13, are composed only by a
single gaussian and have been constructed to change only the behavior
at $r=0$.  The other correlations, shown in the lower panel of
Fig. 13, have the same behavior at $r=0$ but have different
overshooting properties at intermediate distances.  With these
correlations we calculated the $^{16}$O responses at q=400 MeV/$c$
using the S3 set of single particle wave functions. The results of
these calculations are summarized in Fig. 14 where the
$\Delta R_{L,T}$ defined in Eq. (\ref{eq:deltaLT}) are shown as
functions of the excitation energy.

\begin{table}
\caption{Parameters used for the correlation given in Eq. (\ref{gcorr}).}
\label{tab:gcorr}
\begin{center}
\begin{tabular}{lcccccc}
\hline
                & g1   & g2 &  g3  & g4   & g5 & g6  \\
\hline 
  A              & 0.8 & 0.6 & 0.4 & 0.8  & 0.8  & 0.8   \\
  B [fm$^{-2}$]  & 2.2 & 2.0 & 1.9 & 2.2  & 2.2  & 2.2   \\
  C              & 0.0 & 0.0 & 0.0 & 0.04 & 0.04 & 0.04  \\
  D  [fm$^{-2}$] & 0.0 & 0.0 & 0.0 & 3.0  & 3.0  & 3.0   \\
  r$_0$ [fm]     & 0.0 & 0.0 & 0.0 & 1.5  & 2.0  & 2.5   \\
\hline
\end{tabular}
\end{center}
\end{table}

\begin{figure}
\begin{center}
\hspace*{-2.0 cm}
\leavevmode
\epsfysize = 350pt
\epsfbox[70 200 500 650]{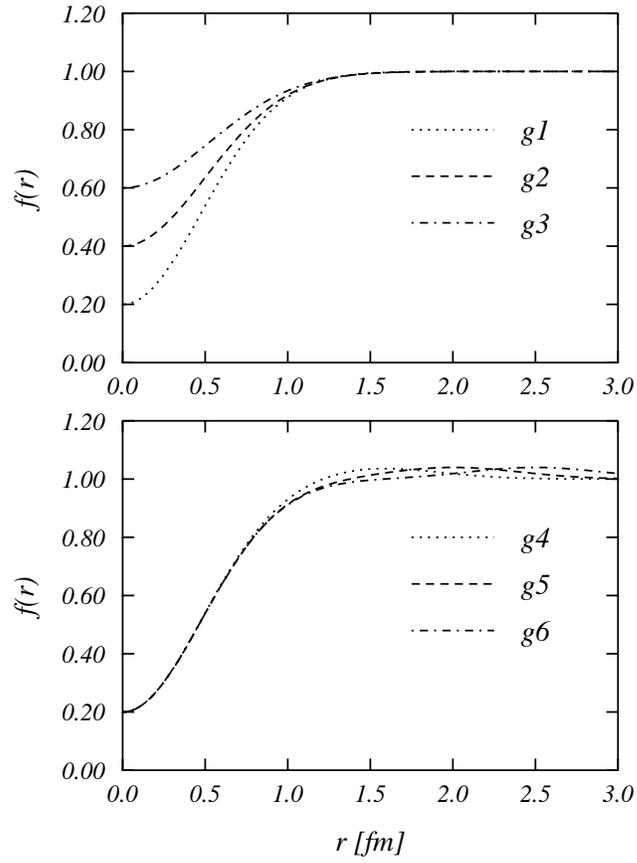}
\end{center}
\vspace{-.5cm}
\caption{\small Correlation obtained with the expression
(\protect\ref{gcorr}) for the various parameterizations given in
Tab. \protect\ref{tab:gcorr}. }
\label{fig:gcorr}
\end{figure}

\begin{figure}
\begin{center}
\hspace*{-.5cm}
\leavevmode
\epsfysize = 300pt
\epsfbox[70 200 500 650]{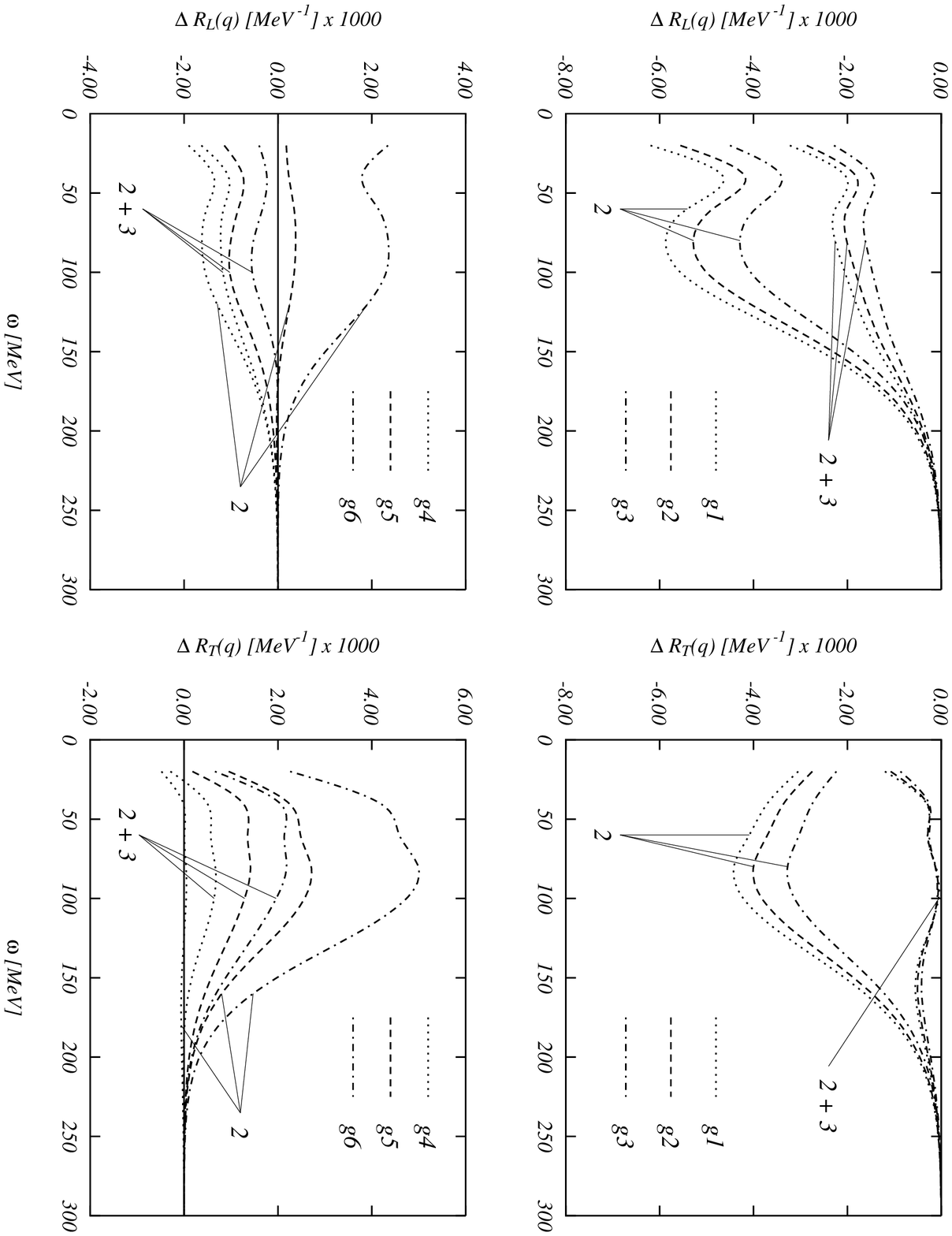}
\end{center}
\vspace{3cm}
\caption{\small Differences between correlated and uncorrelated
$^{16}$O responses at q=400 MeV/$c$ obtained with the S3 single
particle wave functions for the correlations of
Tab. \protect\ref{tab:gcorr}.  The left (right) panels refer to the
longitudinal (transverse) responses.  The curves labelled 2 indicate
the results obtained with only the two-point diagrams of
Fig. \protect\ref{fig:diag1p1h}. The other curves have been obtained
including all the diagrams.}
\label{fig:gdiff}
\end{figure}

The two upper panels of the figure show the effect produced by
changing the correlation depth at $r=0$. The global effect is a
reduction of the MF responses, more evident in the longitudinal
one. The reduction is larger for deeper
correlations.  The other two panels of the figure show the effects of
the overshooting. There is a slight depletion of the longitudinal MF
responses and an enhancement of the transverse ones. The effects are
magnified by shifting the overshooting at larger distances.  This
figure clearly shows the role of the three-point diagrams which always
reduce the effects of the correlations.

To compare of our results with experimental data it is necessary to
consider the effects of the rescattering of the emitted
nucleon with the residual nucleus \cite{ama00}. 
We describe these Final State Interactions (FSI) effects with
the method developed in Refs. \cite{smi88,co88}.
The method shows that, under certain approximations, it is possible to
consider the FSI by folding the bare responses with a smoothing
function: 
\begin{equation}
R^{\rm FSI} (q , \omega) = 
\frac{M^*}{M}
\int_0^\infty dE \, R( q , E)
\left[ \rho \left( E,\frac{M^*}{M} \omega \right) + 
       \rho \left( E, -\frac{M^*}{M} \omega \right) \right] \, ,
\label{rfsi}
\end{equation}
where we have indicated with $M$ and $M^*$ the free and
effective nucleon masses respectively, and 
where the function $\rho (E, \omega)$ describing the effects of FSI
is: 
\begin{equation}
\rho (E, \omega) = 
\displaystyle{\frac{1}{2 \pi} \frac{\Gamma (\omega)}
{ \left[ E - \omega - \Delta (\omega) \right]^2 + 
  \left[ \Gamma (\omega) /2 \right]^2}} \, .
\label{confun}
\end{equation}
The $\Delta (\omega)$ and $\Gamma (\omega)$ functions
of the above equation are connected by a dispersion relation:
\begin{equation}
\Delta (\omega) = \frac{1}{2 \pi} {\cal P} \int_{- \infty}^{+ \infty}
 d \omega ' \frac{\Gamma(\omega ')}{\omega ' - \omega} \, ,
\end{equation}
where ${\cal P}$ indicates the principal value of the integral.
The function $\Gamma( \omega)$ is  
related to the imaginary part of the single particle self-energy.
We have obtained $\Gamma (\omega)$ as average of the single
particle energy width $\gamma(\omega)$:
\begin{equation}
\Gamma (\omega) = \frac{1}{\omega} \int^\infty_0 d \epsilon 
\left[  \gamma(\epsilon+\omega) + \gamma(\epsilon - \omega) \right] \, .
\end{equation}
The single particle energy has not been calculated with a
microscopic model, but, following Ref.\cite{mah82}, has been
parametrized as follows to reproduce empirical values:
\begin{equation}
 \gamma ( \epsilon) = a \cdot  \frac{\epsilon^2}{\epsilon^2 + b^2}
        h(\epsilon) \, ,
\label{spgam}
\end{equation}
with $a=11.5$ and $b=18$. 

The effective nucleon mass takes into account non-locality effects of
the mean-field which cannot be neglected. We determined the
value of the effective nucleon mass by using the expressions 
proposed in the polarization potential model of Ref. \cite{pin88}: 
\begin{equation}
   \frac{M^*}{M} = \frac{1}{1+2M \Delta U/(q^2+q_0^2)} \, ,
\label{emass}
\end{equation}
with 
\begin{equation}
         q_0^2 =\frac{2M \Delta U}{M/M^*_0 -1} \, , 
\end{equation}
$\Delta U$=50 MeV and $M^*_0 =0.8M$.

\begin{figure}[p]
\begin{center}
\hspace*{-2.0 cm}
\leavevmode
\epsfysize = 300pt
\epsfbox[70 200 500 650]{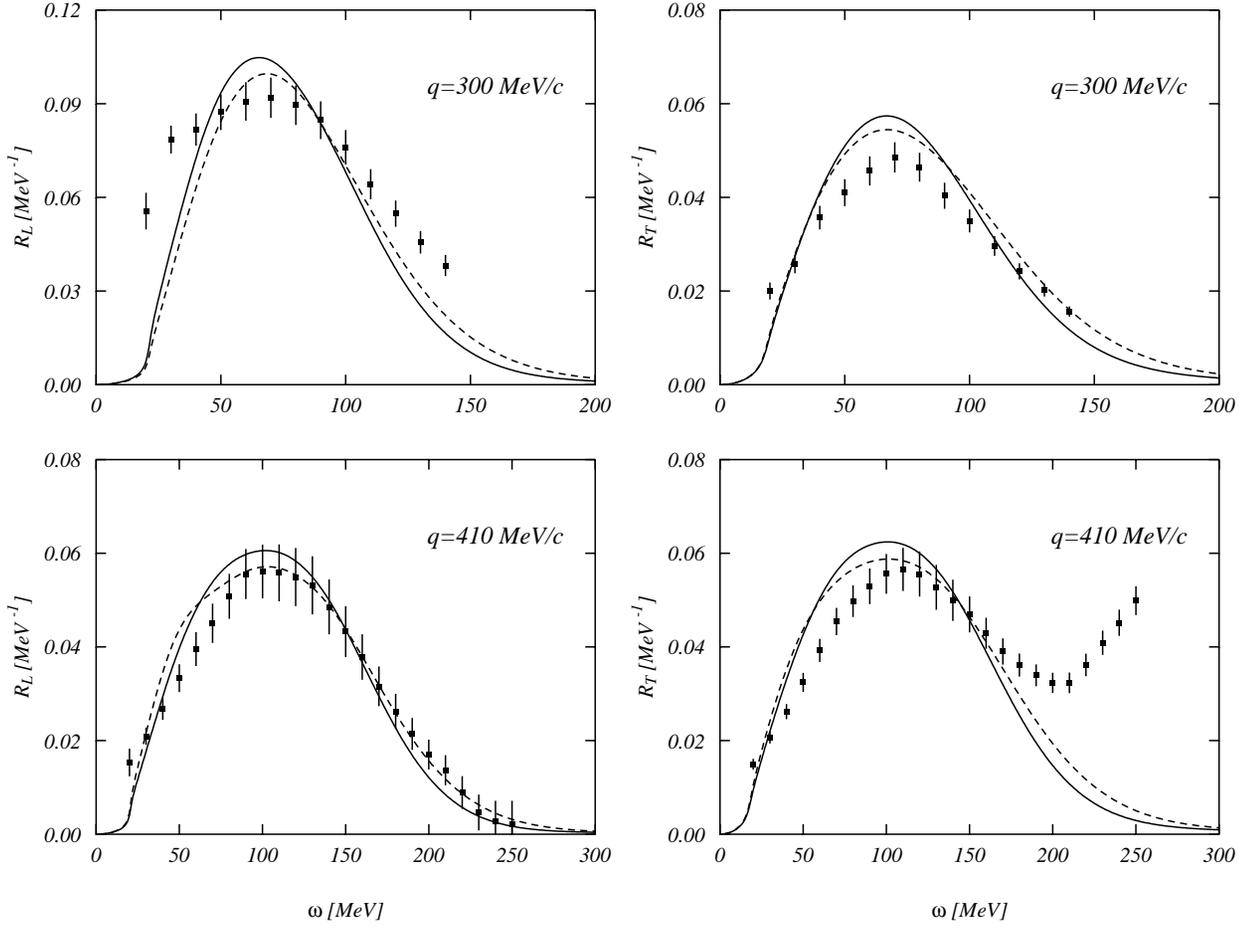}
\end{center}
\vspace{3.0cm}
\caption{\small $^{40}$Ca longitudinal and transverse responses
compared with the data of Ref. \protect\cite{wil97}. The full lines
are the results obtained with the V8 input while the dashed lines have
been obtained with the S3 input.}
\label{fig:ca40exp}
\end{figure}

In Fig. 15 we compare our $^{40}$Ca longitudinal and
transverse responses with the data of Refs. \cite{yat93} and
\cite{wil97}.  Dashed (full) lines show the results of the S3 (V8)
calculations. Both results are in a reasonable agreement with the data.
The differences between the two calculations are mainly produced by
the differences in the mean-field basis rather than in the correlation
functions.

\begin{figure}[p]
\begin{center}
\hspace*{-2.0cm}
\leavevmode
\epsfysize = 300pt
\epsfbox[70 200 500 650]{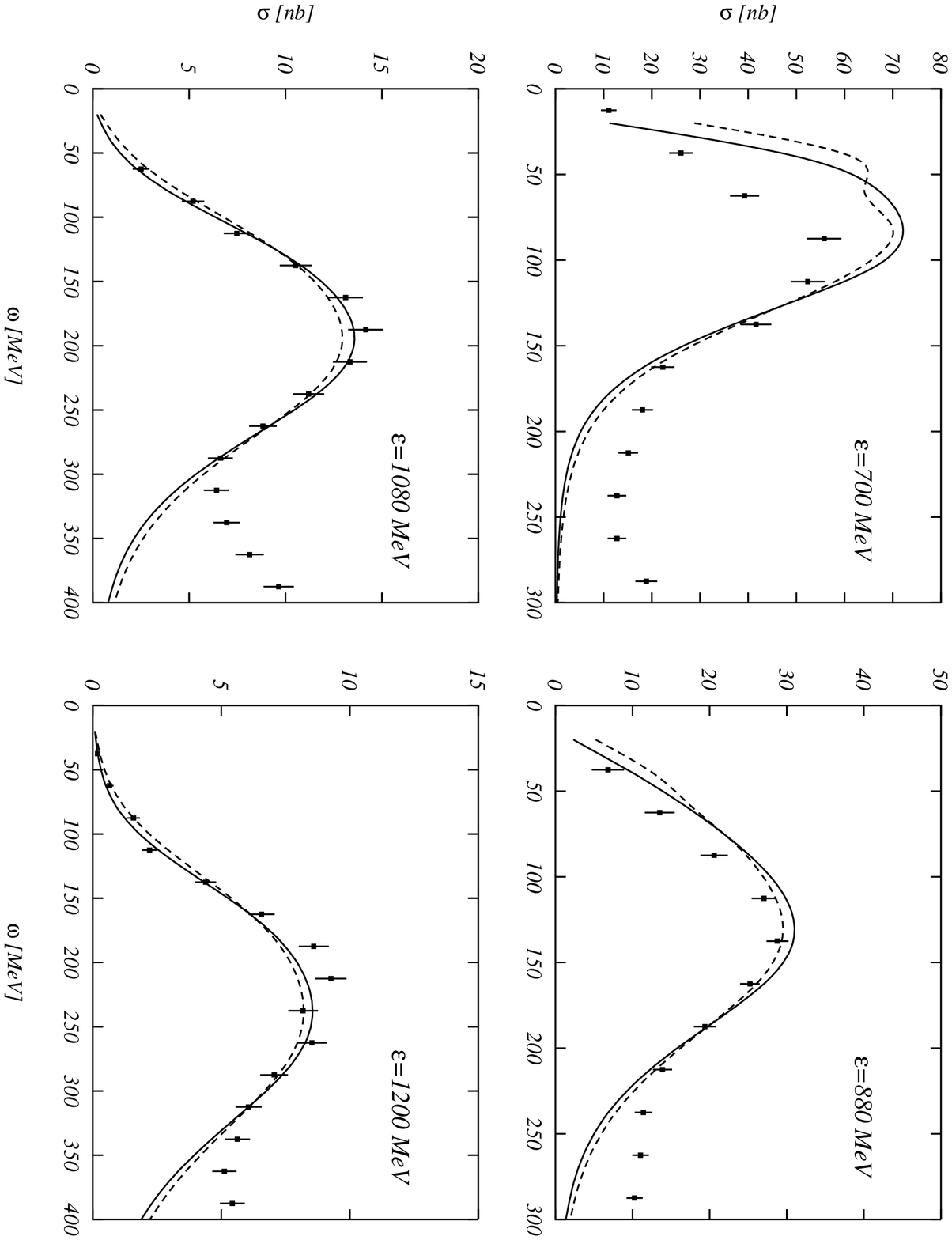}
\end{center}
\vspace{3.0cm}
\caption{\small Electron scattering inclusive cross sections on
$^{16}$O for fixed scattering angle, $\theta=32^{\rm o}$, and various
values of the electron energy. The experimental data are taken from
Ref. \protect\cite{ang96}. The full and dashed lines show the results
obtained with V8 and S3 inputs respectively. }
\label{fig:o16exp}
\end{figure}

For the nucleus $^{16}$O,  Rosenbluth separated
data for longitudinal and transverse responses are not available. 
In Fig. 16 we compare our results with the 
cross section data taken at Frascati \cite{ang96}. 
These data have been measured for fixed values of the incident electron 
energy and of the scattering angle and, therefore, the value of the
momentum transfer changes at every excitation energy.
Since the data have been taken at the scattering angle of 32$^{\rm o}$,
the average momentum value for the cross section at $\epsilon$=700 MeV
is about 400 MeV/$c$ and that at $\epsilon$=1200 MeV is about 650 
MeV/$c$. 

The folding method used to consider FSI is thought for fixed
values of the momentum transfer. Every cross section point has been
calculated by fixing the value of $q$, calculating the response on a
wide energy range (up to 400 MeV), folding the response with
Ed. (\ref{rfsi}), and extracting from this output the single
result at the appropriate excitation energy value. The responses
calculated in this way have been inserted in Eq. (\ref{cross}).

Also in Fig. 16
the full (dashed) lines show the results obtained with the 
V8 (S3) input. The agreement with the data is satisfactory, 
especially for the cross sections measured at higher energies.
The differences between the S3 and V8 results are even
smaller than in the case of $^{40}$Ca and the data cannot 
discriminate between them.

\subsection{Two-nucleon emission}
The study of the two-nucleon emission responses is limited by the
heavy computational load. For this reason we have restricted our
investigation to the $^{16}$O nucleus only.

For these calculations
we used the same configuration space considered in the one-nucleon
emission calculations.
This includes all the protons and neutrons waves with  $l\leq 14$. 
Within this configuration space all the excitation multipoles compatible 
with angular momentum coupling of the 2p-2h excitation pairs have been
considered (see Eq. (\ref{phi2p2h})).

We found that
the contribution of the various partial waves becomes smaller the
higher is the angular momentum value. Specifically, we have seen that
for $q \sim 600$ MeV/$c$ the contribution of
partial waves with $l>4$ is about the 2\% of the total value at an
excitation energy of 100 MeV and becomes about the 40\% 
at the excitation energy of 300 MeV. 
These percentiles remain roughly constant in the momentum transfer range
between 400 and 600 MeV/$c$. 
In order to reduce the computational
time we performed a complete calculation for the partial waves with
$l \leq 4$ while we made a statistical evaluation of the contribution of
the higher partial waves. This statistical evaluation consist in
calculating a randomly chosen 10\% of the 2p-2h matrix elements,
and in rescaling the obtained results proportionally to the total number
of the matrix elements to be calculated. 
This simplification reduced by 2/3 the computational time. 
We have tested the reliability of this simplifying technique by
calculating all the 2p-2h matrix elements  at
the excitation energies of 100, 200 and 300 MeV. 
In all the cases investigated we found differences
with respect to the simplified calculations smaller than  1\%.

The calculations of the two-nucleon emission responses
have been done for both correlation functions at 
excitation energies of 100, 140, 200, 240 and 300 MeV 
for the kinematical
conditions used to evaluate the inclusive cross sections of
Fig. 16 at $\epsilon$ = 700 and 1080 MeV. 
The values of the responses obtained in these calculations 
are shown in Fig. 17 by the squares (S3) and 
by the circles (V8).

\begin{figure}
\begin{center}
\hspace*{-2.0 cm}
\leavevmode
\epsfysize = 300pt
\epsfbox[70 200 500 650]{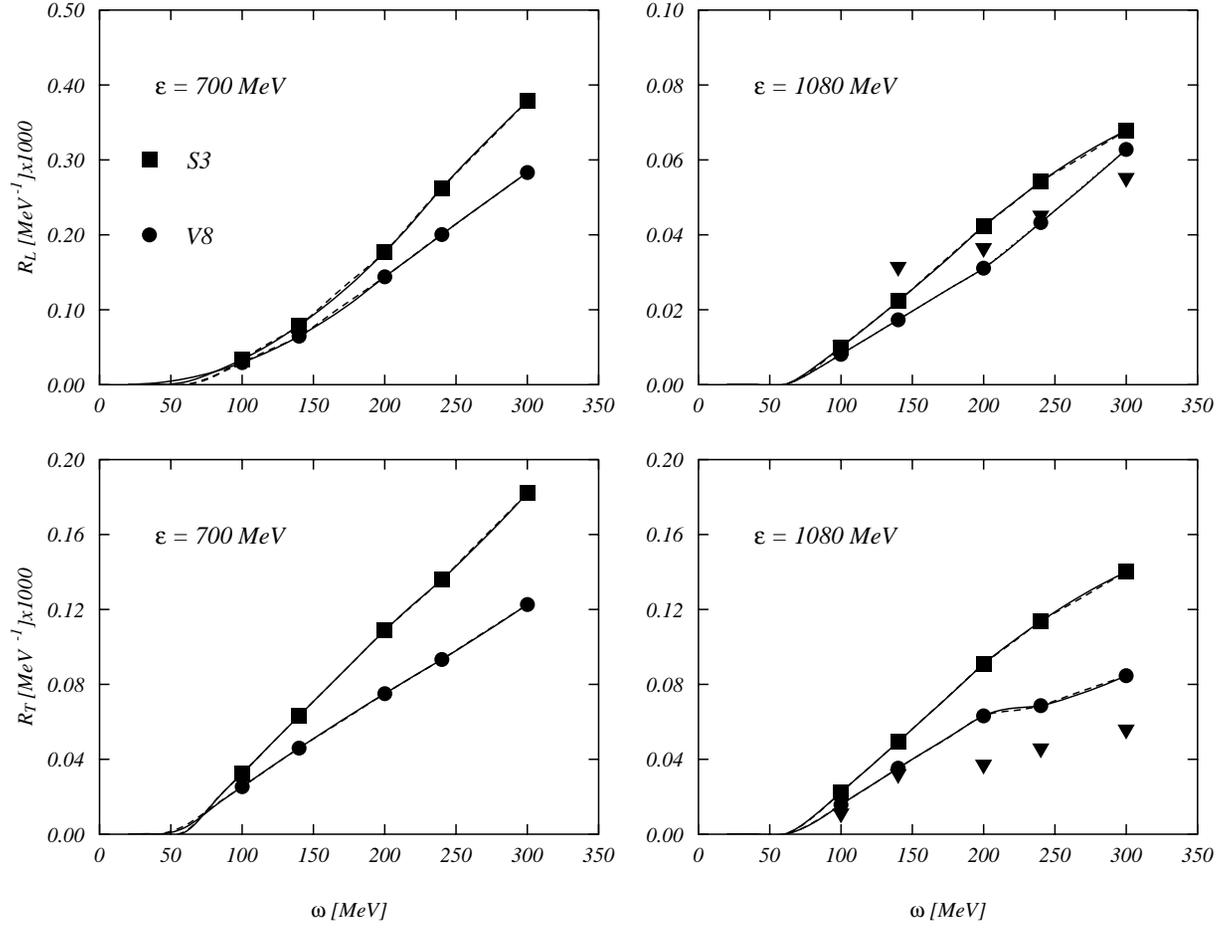}
\end{center}
\vspace{3.0cm}
\caption{\small The symbols show the calculated values of the
longitudinal and transverse two-nucleon emission responses in
$^{16}$O. Squares and circles show the results obtained with the V8
and S3 input respectively. The triangles have been calculated using
the V8 correlation and the S3 mean-field wave functions. The various
lines have been obtained by linear (dashed lines) and quadratic (full
lines) interpolation of the points.}
\label{fig:2p2hpoints}
\end{figure}

A direct comparison with the results of Ref. \cite{co98} is not
straightforward,
since the calculations of that reference have been done on the
$^{12}$C nucleus, and obviously, with different inputs. Considering
these differences we can say that the order of magnitude of the two
calculations is the same. In the two calculations it is however
remarkably different the contribution of the three-point diagrams. 
We found a maximum
contribution to $R_L$ of about 8\% and even smaller, 1\%, for
the transverse response. In the longitudinal response the interference
between two- and three-point diagrams is always destructive, therefore the
final response is always smaller than that of the two-point diagrams
alone. The contradiction with the findings of
Ref. \cite{co98}, where the three-point diagrams were giving an important
contribution to the response, was produced by a computer
error we found in our previous calculations.

The results of  Fig. 17 show a 
large sensitivity to the choice of the correlation.
This is not surprising since in these calculations the two-nucleon
emission is produced only by means of the short range correlation. 
We have also verified the sensitivity to the choice of the mean-field by
calculating responses with V8 correlation and S3 mean-field for
$\epsilon$=1080 MeV. The
results obtained in this way are shown in Fig. 17 by
the triangles. The changes of the mean-field wave functions affect
considerably the final result. 

The two-nucleon emission responses have been used to calculate the
inclusive cross section. The total responses have been obtained by
adding  one- and two-nucleon emission responses. 
To make this sum we
have interpolated the results of Fig. 17 in order to
obtain the values of the 2p-2h responses at the desired excitation
energies. We have used linear and  quadratic interpolations and their
results are shown in Fig. 17 by the dashed and
full lines respectively. These interpolations as well as 
more sophisticated interpolation techniques we have also used produce
results which are the same within the accuracy of the calculations. 
For the evaluation of the cross section we used the results of the 
quadratic interpolation.

The comparison of the results of Fig. 17
with those of Figs. 6 and 7 shows that 
the 2p-2h responses at the peak energy are two orders of
magnitude smaller than the 1p-1h ones. 
The contribution of two-nucleon emission becomes relevant only at
higher energies, in the dip region. As example,
we show in Fig. 18 the contribution of the
various V8 responses in the dip region for the two kinematical cases
we have considered.  
The dashed lines represent the tail of the 1p-1h responses
while the dashed-dotted lines the contribution of the 2p-2h
responses. 
The full lines are the sum of the two. 
It is interesting to notice the
different behavior of the responses for the cases at 700 and 1080
MeV. In the first case, for excitation energies above 220 MeV
the 2p-2h responses become larger than the 1p-1h ones. For
$\epsilon$=1080 MeV the 1p-1h responses remain always larger than the
2p-2h ones. This different behavior is produced by the larger value
of the momentum transverse in this last case which shifts the 1p-1h
responses to higher energies. For this reason the tail of this
response has not yet died out as in the 700 MeV case.

\begin{figure}
\begin{center}
\hspace*{-2.0cm}
\leavevmode
\epsfysize = 300pt
\epsfbox[70 200 500 650]{2respv8.ps}
\end{center}
\vspace{3.0cm}
\caption{\small Longitudinal and transverse responses of $^{16}$O in
the dip region calculated with V8 input.  The dashed lines show the
1p-1h responses, the dashed dotted lines the 2p-2h responses and the
full lines show the total responses.}
\label{fig:2p2hresp}
\end{figure}

\begin{figure}
\begin{center}
\hspace*{-2.0cm}
\leavevmode
\epsfysize = 300pt
\epsfbox[70 200 500 650]{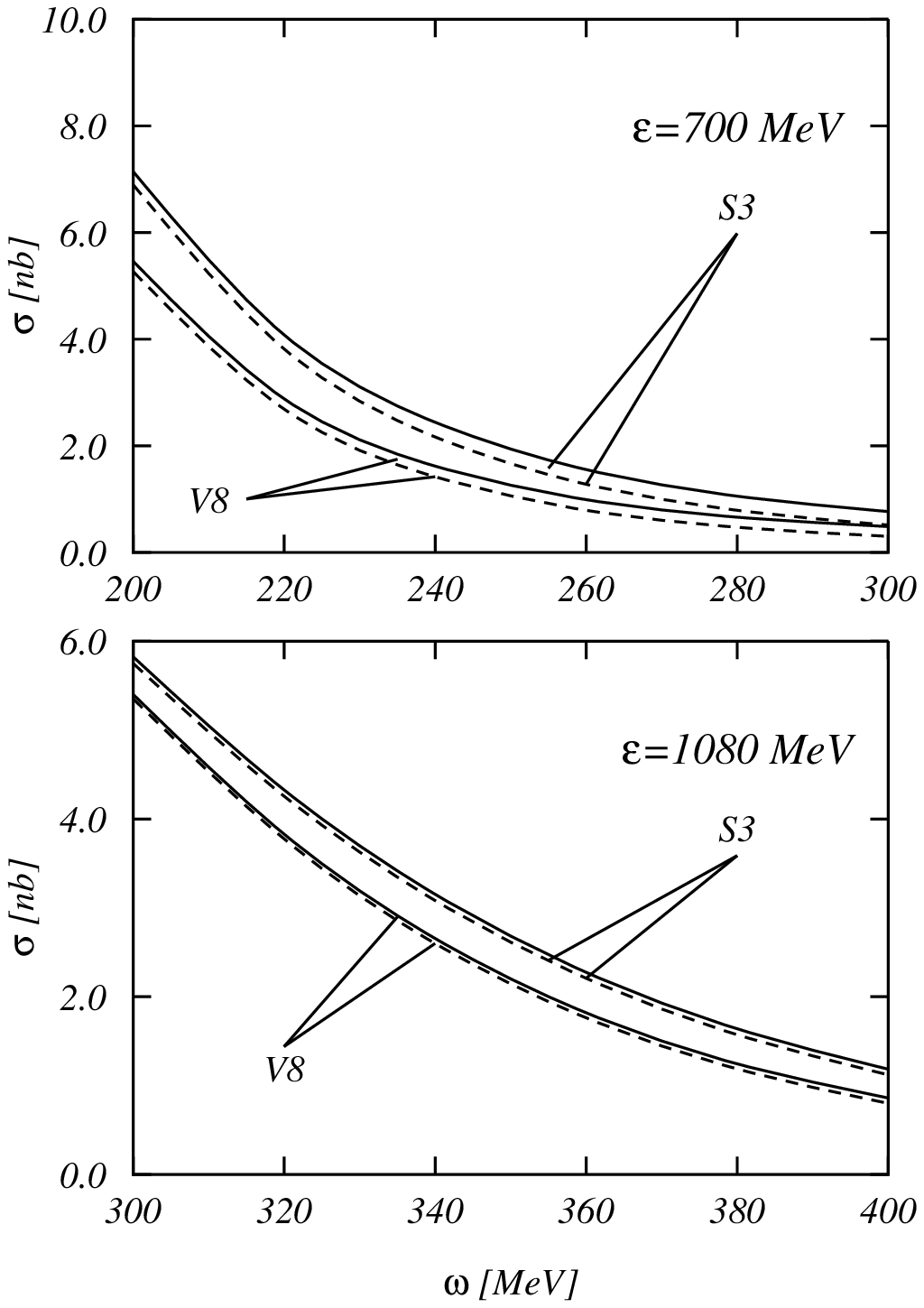}
\end{center}
\caption{\small Inclusive cross sections on $^{16}$O for two values of
the electron incoming energies and a scattering angle of 32$^0$. The
full lines have been obtained including both 1p-1h and 2p-2h
contribution while the dashed lines show the 1p-1h one. }
\label{fig:2p2hcross}
\end{figure}

The calculation of the cross section requires the inclusion of the FSI
effects. We have treated them with the folding model previously
described.  There are ambiguities in our procedure of including the
FSI interaction. We could apply the folding procedure to the total
responses or separately to the 1p-1h and 2p-2h responses and then sum
the two folded responses. The two procedures gave results differing
only by few percent. We show in Fig. 19 the tail of the
calculated cross sections. The dashed lines present the pure 1p-1h
results while the full lines have been obtained by including also the
2p-2h contribution.  The effects of the 2p-2h responses which appeared
in Fig.18 are washed out by the inclusion of the FSI,
whose effects are much larger than those produced by the short range
correlations.  The inclusion of two-nucleon emission responses does
not modify the comparison with the experimental data shown in Fig. 16.

\section{Conclusions}
\label{sect:concl}
In this paper we have presented a model to evaluate the effects of the
short-range correlations in the excitation of the one-body nuclear
responses.
Our model takes into account both two- and three-body diagrams
where the short-range correlations appear at the first order. 
We have shown that both kinds of diagrams are necessary
to have properly normalized wave functions.

As input of our calculations we have used scalar (Jastrow)
correlation functions and single 
particle wave functions fixed in Refs. \cite{ari96} and \cite{fab00} 
by a minimization of the nuclear ground state energy.

The model has been applied to evaluate the inclusive electromagnetic
longitudinal and transverse responses
in the quasi-elastic region for the $^{16}$O and $^{40}$Ca nuclei.
We have considered both one- and two-nucleon emission channels.
The longitudinal response has been calculated using the charge
operator. In the transverse response we have considered the
magnetization current and we have included the convection current only
in the uncorrelated diagram.

Our results show that
the effect of the correlations on the one-particle emission responses
is small,
certainly within the range of uncertainty related to the
different choices of the electromagnetic nucleon form factors.
In spite of this, we made a detailed investigation of short-range
correlations effects in our model because they can be relevant in
exclusive processes.

The three-point diagrams, usually neglected in the literature,
interfere destructively with the two-point diagrams. For this reason
any effect produced by the two-point diagrams is always reduced
when the three-point ones are included. In certain situations this
reduction is of the order of 80\%.  

Our calculations show that the correlations act differently on 
longitudinal and transverse responses. Specifically, 
we found that the longitudinal responses are lowered while 
the transverse ones are enhanced. This happens because some of
the three-point diagrams do not contribute in the longitudinal
response.   

We have calculated the inclusive cross sections in  $^{16}$O to
compare our results with the data of Ref. \cite{ang96}. 
In $^{40}$Ca the comparison has been done directly with the empirical
responses  of Ref. \cite{wil97}.
In both cases the agreement with the data is quite satisfactory 
independently of the inclusion of the short-range correlations.

The two-nucleon emission responses have been calculated only for the 
$^{16}$O nucleus. Because of the heavy computational load we have
simplified the calculation by evaluating the most important matrix
elements and providing a statistical estimation of the others.

As in the one-nucleon emission case, also in the two-nucleon
emission responses, the two- and three-point diagrams
produce effects with opposite sign. However, in this case we found 
that the contribution of the three-point diagrams is at most of the
order of a few percent of that of the two-point diagrams.

In the quasi-elastic region, our calculations show that 
the 2p-2h responses are two order of magnitude smaller than the 1p-1h
ones.  

Summarizing our results, we should say that a proper normalization of
the many body wave function in a first order expansion model for the
correlations, requires the inclusion of the three-body terms. The
contribution of these terms always reduces the effect produced by
the two-point diagrams. 
This effect is not negligible in the one-nucleon emission responses, 
but it becomes rather small in the two-nucleon emission. 

The use of different correlations, together with different set of
single particle wave functions, has clearly shown the sensitivity of
the results to both parts of the input of the calculations. This is
true not only in the 1p-1h responses, where the result was expected
since they are dominated by the uncorrelated terms, but also in the
2p-2h responses, where the uncorrelated term is not present. 
Physically meaningful results can be obtained only if single particle
wave functions and short range correlations are linked through the
nuclear hamiltonian.

In our calculations the presence of short-range correlations 
does not produce remarkable effects.
There are other effects beyond the pure mean-field model, such as
final state interactions and meson exchange currents \cite{ama94},
which are more important.
We should, however, point out that we have used only scalar correlation
functions. There
are indications that state dependent correlations together with
MEC \cite{fab97} could produce sizable effects.

The next step of our work will be to apply the model to the
description of exclusive electron scattering experiments, (e,e'N) and
(e,e'2N), where one expects that the effects of the short-range
correlations could be disentangled.

\section*{Acknowledgments}
We thank M. Anghinolfi and C.F. Williamson for providing us with the
experimental data and J. E. Amaro, F. Arias de Saavedra, A. Fabrocini
and S.M. Rashad for useful discussions.
G.C. thanks the University of Granada for the hospitality during his
visit. 
This work has been partially supported by the CYCIT-INFN agreement and
by the DGES (PB98-1367) and the Junta de Andaluc\'{\i}a (FQM225).

\newpage

\section*{Appendix A. One-particle emission matrix elements}

In this appendix we give the expressions of the matrix elements
involved in the 
one-nucleon emission part of the longitudinal and transverse
responses. 
As discusses in Sect. \ref{sect:emexcitations}, only the charge
operator contributes to the longitudinal response, 
while both the electric and magnetic parts of the magnetization
current, Eqs. (\ref{tejm},\ref{tmjm}), contribute to transverse response. 
The convection current is considered in the one-point diagram only.  

We begin with the Coulomb (charge) matrix elements for which we have:
\begin{eqnarray}
M^{(\rm 1p1h)}_1 & = & G_E^p(q^2) \, \delta_{t_p,t_h } \,
\gamma^J_{ph} \, {\cal I}_J^{[ph]}(q) \, , \\[5mm]
M^{(\rm 1p1h)}_{2.1} & = & G_E^p(q^2) \, 
\delta_{t_p,t_h } \, 4\pi \, \gamma^J_{ph} \, 
{\cal I}_{J0}^{[ph];[\rho]}(q) \, , \\[.5cm]
M^{(\rm 1p1h)}_{2.2} & = & G_E^p(q^2) \, \delta_{t_p,t_h } \,
\gamma^J_{ph} \, \sum_{i L} \, \delta_{t_i,t_p} \,
\xi(l_i+l_h+L) \, \widehat{j}_i^2 \\ \nonumber &&
{\threej{j_i}{j_h}{L}{\half}{-\half}{0}}^2 \, 
{\cal I}_{JL}^{[pi];[ih]}(q) \, , \\[.5cm]
M^{(\rm 1p1h)}_{2.3} & = & \delta_{t_p,t_h } \, 4\pi \, \gamma^J_{ph} \,
\frac{1}{\widehat{J}^2} \, {\cal I}_{JJ}^{[\rho_{\rm ch}];[ph]}(q)
\, , \\[.5cm]
M^{(\rm 1p1h)}_{2.4} & = & G_E^p(q^2) \, \delta_{t_p,t_h } \,
\gamma^J_{ph} \,  \sum_{i L} \, \delta_{t_i,t_p} \,
\xi(l_p+l_i+L) \, \widehat{j}_i^2 \\ \nonumber
&&{\threej{j_p}{j_i}{L}{\half}{-\half}{0}}^2 \, 
{\cal I}_{JL}^{[ih];[pi]}(q) \, , \\[.5cm]
M^{(\rm 1p1h)}_{3.1} & = & G_E^p(q^2) \, \delta_{t_p,t_h } \,
\gamma^J_{ph} \, \sum_{i k L} \, \delta_{t_i,t_p} \,
\delta_{t_k,t_p} \, \delta_{j_i,j_p} \, \xi(l_p+l_k+L) \, \xi(l_p+l_i)
\, \widehat{j}_k^2  \\ \nonumber & ~~ &
{\threej{j_p}{j_k}{L}{\half}{-\half}{0}}^2 \,
{\cal I}_{J}^{[ih]}(q) \,\, {\cal J}_{L}^{[pk];[ki]} \, , \\[.5cm]
M^{(\rm 1p1h)}_{3.2} & = & G_E^p(q^2) \, \delta_{t_p,t_h } \, 4\pi \,
\gamma^J_{ph} \, \sum_{i} \, \delta_{t_i,t_p} \, \delta_{j_i,j_p} \,
\xi(l_p+l_i) \,\, {\cal I}_{J}^{[ih]}(q) \,\, {\cal
J}_{0}^{[pi];[\rho]} 
\, , \\[.5cm]
M^{(\rm 1p1h)}_{3.3} & = & G_E^p(q^2) \, \delta_{t_p,t_h } \,
\gamma^J_{ph} \, \sum_{i k L} \, \delta_{t_i,t_p} \,
\delta_{t_k,t_p} \, \delta_{j_i,j_h} \, \xi(l_k+l_h+L) \, \xi(l_i+l_h)
\, \widehat{j}_k^2 \\ \nonumber 
& ~~ & {\threej{j_k}{j_h}{L}{\half}{-\half}{0}}^2 \,
{\cal I}_{J}^{[pi]}(q) \,\, {\cal J}_{L}^{[kh];[ik]} \, , \\[.5cm]
M^{(\rm 1p1h)}_{3.4} & = & G_E^p(q^2) \, \delta_{t_p,t_h } \, 4\pi \, 
\gamma^J_{ph} \, \sum_{i} \, \delta_{t_i,t_p} \, \delta_{j_i,j_h} \,
\xi(l_i+l_h) \,\, {\cal I}_{J}^{[pi]}(q) \,\, {\cal J}_{0}^{[ih];[\rho]}
\, , \\[.5cm]
M^{(\rm 1p1h)}_{3.5} & = & G_E^p(q^2) \, \delta_{t_p,t_h } \, 4\pi
 \, \sum_{i k L} \, \delta_{t_i,t_p} \, \delta_{t_k,t_p}
 \, (-1)^{j_i+j_k+J+L} \, \frac{1}{\widehat{L}^2} \, \gamma^J_{ik} \,
 \gamma^L_{pi} \, \gamma^L_{kh} \, \\
 \nonumber & ~~ & \sixj{j_p}{j_h}{J}{j_k}{j_i}{L} \,\,
 {\cal I}_{J}^{[ik]}(q) \,\, {\cal J}_{L}^{[pi];[kh]} \, , \\[.5cm]
M^{(\rm 1p1h)}_{3.6} & = & \delta_{t_p,t_h } \, 4\pi \, 
\frac{1}{\widehat{J}^4} \, \gamma^J_{ph} \, 
\sum_{i k} \, \delta_{t_i,t_k} \, G_E^i(q^2) \, \, 
(\gamma^J_{ik})^2 {\cal
I}_{J}^{[ik]}(q) \, {\cal J}_{J}^{[ph];[ki]} \, .
\end{eqnarray}

In the previous equations we have indicated with $G_E^\alpha(q^2)$
the electric form factor of the nucleon in the single particle state 
$\alpha$, and we have defined the integrals
\begin{eqnarray}
{\cal I}_J^{[\alpha \beta]}(q) & = & \int {\rm d}r \, r^2 \,
j_J(qr) \, R_\alpha^*(r) \, R_\beta(r) \, , \\ 
{\cal I}_{JL}^{[\alpha \beta];[\gamma \delta]}(q) & = &
\int {\rm d}r_1 \, r_1^2 \, j_J(qr) \, 
R^*_\alpha(r_1) \, R_\beta(r_1) \, {\cal H}_L^{[\gamma \delta]}(r_1) \, ,
\\
{\cal I}_{J0}^{[\alpha \beta];[\rho]}(q) & = &
\int {\rm d}r_1 \, r_1^2 \, j_J(qr) \, 
R^*_\alpha(r_1) \, R_\beta(r_1) \, {\cal H}_0^{[\rho]}(r_1) \, , \\
{\cal I}_{JJ}^{[\rho_{\rm ch}];[\gamma \delta]}(q) & = & 
\int {\rm d}r_1 \, r_1^2 \, j_J(qr) \, \rho_{\rm ch}(r_1)
\, {\cal H}_J^{[\gamma \delta]}(r_1) \, ,
\end{eqnarray}
where $j_J(x)$ is a spherical Bessel function,
${\cal H}_L^{[\gamma \delta]}(r)$ is given by Eq. (\ref{calHint}),
${\cal H}_0^{[\rho]}(r)$ is given by Eq. (\ref{calH0rho}), and
\begin{equation}
\rho_{\rm ch}(r) \, = \, G_E^{\rm prot}(q^2) \, \rho_{\rm prot}(r) \, +
\, G_E^{\rm neut}(q^2) \, \rho_{\rm neut}(r) \, , 
\end{equation}
with $\rho_{\rm prot}(r)$ and $\rho_{\rm neut}(r)$ the proton and neutron 
densities, respectively. Finally, the integrals ${\cal J}$ are
defined in Eqs. (\ref{calJint},\ref{calJ0int}). 

The matrix elements of the transverse electric part of the
magnetization current are given by:
\begin{eqnarray}
T_{1;m}^{{\rm E}(\rm 1p1h)} 1 & = & - q \, \frac{G_M^p(q^2)}{2M_p} \, 
\delta_{t_p,t_h} \,
\gamma^J_{ph} \, \frac{\chi_p-\chi_h}{\sqrt{J(J+1)}} \,
{\cal I}_J^{[ph]}(q) \, , \\[5mm]
T_{2.1;m}^{{\rm E}(\rm 1p1h)} & = & -q \, \frac{G_M^p(q^2)}{2M_p} \, 
\delta_{t_p,t_h } \, 4\pi \, \gamma^J_{ph} \, 
\frac{\chi_p-\chi_h}{\sqrt{J(J+1)}} \, {\cal I}_{J0}^{[ph];[\rho]}(q)
\, , \\[.5cm]
T_{2.2;m}^{{\rm E}(\rm 1p1h)}  & = & q \, \frac{G_M^p(q^2)}{2M_p} \, 
\delta_{t_p,t_h } \,
\sqrt{\frac{3}{\pi}} \, (-1)^{l_p+j_p+J+\frac{1}{2}} \,
\widehat{l}_p \, \widehat{j}_p \, \widehat{j}_h \, \widehat{J} 
\, \sum_{i L} \, \delta_{t_i,t_p} \, \xi(l_i+l_h+L) \,
\widehat{l}_i \, \widehat{j}_i^2 \nonumber \\ 
& ~~ & 
\threej{j_i}{j_h}{L}{\half}{-\half}{0} \, {\cal I}_{JL}^{[pi];[ih]}(q) \,
\sum_\lambda \, \xi(L+J+\lambda) \, \widehat{\lambda}^2 \,
\threej{l_p}{l_i}{\lambda}{0}{0}{0} \\ \nonumber
& ~~ & \sum_K \, \widehat{K}^2 \,
\sixj{L}{K}{J}{j_p}{j_h}{j_i}
\threej{1}{K}{\lambda}{1}{-1}{0}
\threej{K}{J}{L}{1}{-1}{0}
\ninej{l_p}{\half}{j_p}{l_i}{\half}{j_i}{\lambda}{1}{K} \, ,\\[.5cm]
T_{2.3;m}^{{\rm E}(\rm 1p1h)}  & = & 0 \, ,\\[.5cm]
T_{2.4;m}^{{\rm E}(\rm 1p1h)}  & = & q \, \frac{G_M^p(q^2)}{2M_p} \, 
\delta_{t_p,t_h } \,
\sqrt{\frac{3}{\pi}} \, (-1)^{l_p+j_p+j_h} \,
\widehat{l}_h \, \widehat{j}_p \, \widehat{j}_h \, \widehat{J} \,
\sum_{i L} \, \delta_{t_i,t_p} \, (-1)^{j_i-\frac{1}{2}} 
\, \xi(l_p+l_i+L) \nonumber \\ && \widehat{l}_i \, \widehat{j}_i^2 \, 
\threej{j_p}{j_i}{L}{\half}{-\half}{0} \, {\cal I}_{JL}^{[ih];[pi]}(q)
\, \sum_\lambda \, \xi(L+J+\lambda) \, \widehat{\lambda}^2 \,
\threej{l_i}{l_h}{\lambda}{0}{0}{0} \\ \nonumber
& ~~ & 
\sum_K \, (-1)^K \, \widehat{K}^2 \, \sixj{L}{K}{J}{j_h}{j_p}{j_i}
\threej{1}{K}{\lambda}{1}{-1}{0}
\threej{K}{J}{L}{1}{-1}{0} 
\ninej{l_i}{\half}{j_i}{l_h}{\half}{j_h}{\lambda}{1}{K} \, , \\[5mm]
T_{3.1;m}^{{\rm E}(\rm 1p1h)}  & = & - q \, \frac{G_M^p(q^2)}{2M_p} \, 
\delta_{t_p,t_h } \,
\gamma^J_{ph} \, \frac{\chi_p-\chi_h}{\sqrt{J(J+1)}} \,
\sum_{i k L} \, \delta_{t_i,t_p} \, \delta_{t_k,t_p} 
\, \delta_{j_i,j_p} \\ \nonumber &&
\xi(l_p+l_k+L) \, \xi(l_p+l_i) \,  \widehat{j}_k^2 \,
{\threej{j_p}{j_k}{L}{\half}{-\half}{0}}^2 \, {\cal I}_{J}^{[ih]}(q)
\, {\cal J}_{L}^{[pk];[ki]} \, ,\\[.5cm]
T_{3.2;m}^{{\rm E}(\rm 1p1h)}  & = & -q \, \frac{G_M^p(q^2)}{2M_p} \, 
\delta_{t_p,t_h } \, 4\pi \,
\gamma^J_{ph} \, \frac{\chi_p-\chi_h}{\sqrt{J(J+1)}} \,
\sum_{i} \, \delta_{t_i,t_p} \, \delta_{j_i,j_p} \, \xi(l_p+l_i) 
\\ \nonumber &&
{\cal I}_{J}^{[ih]}(q) \, {\cal J}_{0}^{[pi];[\rho]} \, ,\\[.5cm]
T_{3.3;m}^{{\rm E}(\rm 1p1h)}  & = & -q \, \frac{G_M^p(q^2)}{2M_p} \, 
\delta_{t_p,t_h } \,
\gamma^J_{ph} \, \frac{\chi_p-\chi_h}{\sqrt{J(J+1)}} \,
\sum_{i k L} \, \delta_{t_i,t_p} \, \delta_{t_k,t_p} 
\, \delta_{j_i,j_h} \\ \nonumber && \xi(l_k+l_h+L) \, \xi(l_i+l_h) \,
\widehat{j}_k^2 \,
{\threej{j_k}{j_h}{L}{\half}{-\half}{0}}^2 \, {\cal I}_{J}^{[pi]}(q)
\, {\cal J}_{L}^{[kh];[ik]} \, ,\\[.5cm]
T_{3.4;m}^{{\rm E}(\rm 1p1h)}  & = & -q \, \frac{G_M^p(q^2)}{2M_p} \, 
\delta_{t_p,t_h } \, 4\pi \,
\gamma^J_{ph} \, \frac{\chi_p-\chi_h}{\sqrt{J(J+1)}} \,
\sum_{i} \, \delta_{t_i,t_p} \, \delta_{j_i,j_h} \, \xi(l_i+l_h)
\\ \nonumber &&
{\cal I}_{J}^{[pi]}(q) \, {\cal J}_{0}^{[ih];[\rho]} \, ,\\[.5cm]
T_{3.5;m}^{{\rm E}(\rm 1p1h)} & = & -q \, \frac{G_M^p(q^2)}{2M_p} \, 
\delta_{t_p,t_h } \, 4\pi \, \frac{1}{\sqrt{J(J+1)}} \, \sum_{i k L} \, 
\delta_{t_i,t_p} \, \delta_{t_k,t_p} \, (-1)^{j_i+j_k+J+L} \,
\frac{\chi_i-\chi_k}{\widehat{L}^2} 
\\ \nonumber &&
\gamma^J_{ik} \, \gamma^L_{pi} \, \gamma^L_{kh} \,
\sixj{j_p}{j_h}{J}{j_k}{j_i}{L} \, {\cal I}_{J}^{[ik]}(q)
\, {\cal J}_{L}^{[pi];[kh]} \, , \\[.5cm]
T_{3.6;m}^{{\rm E}(\rm 1p1h)} & = & - \delta_{t_p,t_h } \, 4\pi \, 
\frac{1}{\widehat{J}^4 \sqrt{J(J+1)}} \, \gamma^J_{ph} \, 
\sum_{i k} \, \delta_{t_i,t_k} \, q \, \frac{G_M^i(q^2)}{2M_i} \,
(\chi_i-\chi_k) \, (\gamma^J_{ik})^2 \\
\nonumber && {\cal I}_{J}^{[ik]}(q)
\, {\cal J}_{J}^{[ph];[ki]} \, .
\end{eqnarray}

In these equations we have indicated with $G_M^\alpha(q^2)$
the magnetic form factor of the nucleon in 
the single particle state $\alpha$,
and with $M_\alpha$. The symbol $\chi_\alpha$ has been defined as:
\begin{equation}
\chi_\alpha \, = \, (l_\alpha - j_\alpha) \, (2j_\alpha+1) \, .
\end{equation}

For the transverse magnetic part of the
magnetization current the corresponding matrix elements are:
\begin{eqnarray}
T_{1;m}^{{\rm M}(\rm 1p1h)}  & = & -i q \, \frac{G_M^p(q^2)}{2M_p} \, 
\delta_{t_p,t_h } \,
\eta^J_{ph} \, \frac{1}{\widehat{J}^2} \\ \nonumber &&
\sum_{s=\pm1} \, s \, \sqrt{\frac{J+\delta_{s,-1}}{J+\delta_{s,1}}} \,
(\chi_p+\chi_h+sJ+\delta_{s,1})\,
{\cal I}_{J'}^{[ph]}(q) \, ,\\[.5cm]
T_{2.1;m}^{{\rm M}(\rm 1p1h)}  & = & -i q \, \frac{G_M^p(q^2)}{2M_p} \, 
\delta_{t_p,t_h } \, 4\pi \, \eta^J_{ph} \, \frac{1}{\widehat{J}^2} 
\\ \nonumber && \sum_{s=\pm1} \, s \,
\sqrt{\frac{J+\delta_{s,-1}}{J+\delta_{s,1}}} \,
(\chi_p+\chi_h+sJ+\delta_{s,1})\, {\cal I}_{J'0}^{[ph];[\rho]}(q) 
\, , \\[.5cm]
T_{2.2;m}^{{\rm M}(\rm 1p1h)}  & = & -i q \, \frac{G_M^p(q^2)}{2M_p} \, 
\delta_{t_p,t_h } \,
\sqrt{\frac{3}{2\pi}} \, (-1)^{j_h+j_p} \,
\widehat{l}_p \, \widehat{j}_p \, \widehat{j}_h \nonumber \\ 
& ~~ & \sum_{i L} \, \delta_{t_i,t_p} \, (-1)^{j_i+l_i+\frac{1}{2}} \,
\xi(l_i+l_h+L) \, \widehat{l}_i \, \widehat{j}_i^2 \,
\threej{j_i}{L}{j_h}{\half}{0}{-\half} \\  \nonumber 
&& \sum_{s=\pm1} \, s
\,\sqrt{J+\delta_{s,-1}} \, \widehat{J'} \,
{\cal I}_{J'L}^{[pi];[ih]}(q) \,
\sum_\lambda \, \widehat{\lambda}^2 \,
\threej{l_p}{l_i}{\lambda}{0}{0}{0} \, 
\threej{L}{J'}{\lambda}{0}{0}{0} \\ 
 \nonumber  & ~~ & \sum_K \, (-1)^K \, \widehat{K}^2 \,
\sixj{L}{K}{J}{j_p}{j_h}{j_i}
\sixj{L}{J'}{\lambda}{1}{K}{J}
\ninej{l_p}{\half}{j_p}{l_i}{\half}{j_i}{\lambda}{1}{K} \, ,
\\[.5cm]
T_{2.3;m}^{{\rm M}(\rm 1p1h)}  & = & 0 \, ,\\[.5cm]
T_{2.4;m}^{{\rm M}(\rm 1p1h)}  & = & -i q \, \frac{G_M^p(q^2)}{2M_p} \, 
\delta_{t_p,t_h } \,
\sqrt{\frac{3}{2\pi}} \, (-1)^{j_h+\frac{1}{2}} \,
\widehat{l}_h \, \widehat{j}_p \, \widehat{j}_h \nonumber  \\
& ~~ & \sum_{i L} \, \delta_{t_i,t_p} \, (-1)^{l_i} \,
\xi(l_p+l_i+L) \, \widehat{l}_i \, \widehat{j}_i^2 \,
\threej{j_p}{L}{j_i}{\half}{0}{-\half} \\  \nonumber && \sum_{s=\pm1} \, s
\,\sqrt{J+\delta_{s,-1}} \, \widehat{J'} \, 
{\cal I}_{J'L}^{[ih];[pi]}(q) \,
\sum_\lambda \, \widehat{\lambda}^2 \,
\threej{l_i}{l_h}{\lambda}{0}{0}{0} \, 
\threej{L}{J'}{\lambda}{0}{0}{0} \\ 
 \nonumber & ~~ & \sum_K \, \widehat{K}^2 \,
\sixj{L}{K}{J}{j_h}{j_p}{j_i}
\sixj{L}{J'}{\lambda}{1}{K}{J}
\ninej{l_i}{\half}{j_i}{l_h}{\half}{j_h}{\lambda}{1}{K} \, , \\[.5cm] 
T_{3.1;m}^{{\rm M}(\rm 1p1h)}  & = & - iq \, \frac{G_M^p(q^2)}{2M_p} \, 
\delta_{t_p,t_h } \, \eta^J_{ph} \, 
\frac{1}{\widehat{J}^2} \,
\sum_{i k L} \, \delta_{t_i,t_p} \, \delta_{t_k,t_p} 
\, \delta_{j_i,j_p} \nonumber \\ && 
\xi(l_p+l_k+L) \, \xi(l_p+l_i) \, \widehat{j}_k^2 \,
{\threej{j_p}{j_k}{L}{\half}{-\half}{0}}^2 
\, {\cal J}_{L}^{[pk];[ki]} \\ \nonumber & ~~ & \sum_{s=\pm1} \, s \,
\sqrt{\frac{J+\delta_{s,-1}}{J+\delta_{s,1}}} \,
(\chi_p+\chi_h+sJ+\delta_{s,1}) \,
{\cal I}_{J'}^{[ih]}(q) \, ,\\[.5cm]
T_{3.2;m}^{{\rm M}(\rm 1p1h)}  & = & -i q \, \frac{G_M^p(q^2)}{2M_p} \, 
\delta_{t_p,t_h } \, 4\pi \, \eta^J_{ph} \, 
\frac{1}{\widehat{J}^2} \, \sum_{i} \, \delta_{t_i,t_p} \,
\delta_{j_i,j_p} \, \xi(l_p+l_i) \,{\cal J}_{0}^{[pi];[\rho]} \\
\nonumber &&
\sum_{s=\pm1} \, s \,
\sqrt{\frac{J+\delta_{s,-1}}{J+\delta_{s,1}}} \,
(\chi_p+\chi_h+sJ+\delta_{s,1}) \, 
{\cal I}_{J'}^{[ih]}(q)  \, , \\[.5cm]
T_{3.3;m}^{{\rm M}(\rm 1p1h)}  & = & - i q \, \frac{G_M^p(q^2)}{2M_p} \, 
\delta_{t_p,t_h } \, \eta^J_{ph} \, 
\frac{1}{\widehat{J}^2} \, 
\sum_{i k L} \, \delta_{t_i,t_p} \, \delta_{t_k,t_p} 
\, \delta_{j_i,j_h} \nonumber \\ && 
\xi(l_k+l_h+L) \, \xi(l_i+l_h) \, \widehat{j}_k^2 \,
{\threej{j_k}{j_h}{L}{\half}{-\half}{0}}^2 \, 
{\cal J}_{L}^{[kh];[ik]} \\ \nonumber
& ~~ & \nonumber  \sum_{s=\pm1} \, s \,
\sqrt{\frac{J+\delta_{s,-1}}{J+\delta_{s,1}}} \,
(\chi_p+\chi_h+sJ+\delta_{s,1}) \,
{\cal I}_{J'}^{[pi]}(q) \, , \\[.5cm]
T_{3.4;m}^{{\rm M}(\rm 1p1h)}  & = & -i q \, \frac{G_M^p(q^2)}{2M_p} \, 
\delta_{t_p,t_h } \, 4\pi  \, \eta^J_{ph} \, 
\frac{1}{\widehat{J}^2} \, 
\sum_{i} \, \delta_{t_i,t_p} \, \delta_{j_i,j_h} \, \xi(l_i+l_h) \,
 {\cal J}_{0}^{[ih];[\rho]} \\ \nonumber &&
\sum_{s=\pm1} \, s \, \sqrt{\frac{J+\delta_{s,-1}}{J+\delta_{s,1}}} \,
(\chi_p+\chi_h+sJ+\delta_{s,1}) {\cal I}_{J'}^{[pi]}(q) \, , \\[.5cm]
T_{3.5;m}^{{\rm M}(\rm 1p1h)}  & = & -i q \, \frac{G_M^p(q^2)}{2M_p} \, 
\delta_{t_p,t_h } \, 4\pi \, \frac{1}{\widehat{J}^2} \,
\sum_{i k L} \, \delta_{t_i,t_p} \,
\delta_{t_k,t_p} \, (-1)^{j_i+j_k+J+1} \, \eta^J_{ik}
\frac{1}{\widehat{L}^2} \nonumber \\  && 
\gamma^L_{pi} \, \gamma^L_{kh} \, \sixj{j_p}{j_h}{J}{j_k}{j_i}{L} \, 
\threej{j_i}{j_k}{L}{\half}{-\half}{0} \, {\cal J}_{L}^{[pi];[kh]} \\ 
\nonumber && \sum_{s=\pm1} \, s \,
\sqrt{\frac{J+\delta_{s,-1}}{J+\delta_{s,1}}} \,
(\chi_i+\chi_k+sJ+\delta_{s,1}) \,  
{\cal I}_{J'}^{[ik]}(q)
\, , \\[.5cm]
T_{3.6;m}^{{\rm M}(\rm 1p1h)}  & = & 0 \, .
\end{eqnarray}

In the previous equations we have defined
\begin{eqnarray} 
\nonumber \eta^{\lambda}_{\alpha \beta} \, = \, 
\frac{1}{\sqrt{4\pi}} \, (-1)^{j_{\beta}+\lambda-\frac12} \,
\xi(l_\alpha+l_\beta+\lambda+1) \, \widehat{j}_\alpha \widehat{j}_\beta
\widehat{\lambda} \,
\threej{j_\alpha}{j_\beta}{\lambda}{\half}{-\half}{0} 
\end{eqnarray}
and $J'$ stands for $J+s$.

Finally, for the transverse electric and magnetic parts of the convection
current contributing to the one-point diagram we have:
\begin{eqnarray}
T_{1;c}^{{\rm E}(\rm 1p1h)} & = & \frac{1}{q} \, \frac{G_E^p(q^2)}{2M_p} \, 
\delta_{t_p,t_h } \, \gamma^J_{ph} \, \frac{1}{\sqrt{J(J+1)}}
\nonumber \\ & ~~ & \left\{
[(\chi_p-\chi_h)(\chi_p+\chi_h+1)+J(J+1)] \, 
{\cal I}_J^{[\frac1r ph']}(q) \,
+ \right. \\ \nonumber && \left. [(\chi_p-\chi_h)(\chi_p+\chi_h+1)-J(J+1)] \,
{\cal I}_J^{[\frac1r p'h]}(q)
\right\} \, , \\[5mm]
T_{1;c}^{{\rm M}(\rm 1p1h)} & = & -i \, \frac{G_E^p(q^2)}{2M_p} \, 
\delta_{t_p,t_h } \, \eta^J_{ph} \, 
\frac{(\chi_p+\chi_h)(\chi_p+\chi_h+1)-J(J+1)}{\sqrt{J(J+1)}} 
\, {\cal I}_J^{[\frac1r ph]}(q) \, ,
\end{eqnarray}
where we have used the following definitions:
\begin{eqnarray}
{\cal I}_J^{[\frac1r p h']}(q) & = & \int {\rm d}r \, r \,
j_J(qr) \, R_p^*(r) \, \frac{{\rm d}R_h(r)}{{\rm d}r} \, , \\[5mm]  
{\cal I}_J^{[\frac1r p'h]}(q) & = & \int {\rm d}r \, r \,
j_J(qr) \, \frac{{\rm d}R_p^*(r)}{{\rm d}r} \, R_h(r) \, , \\[5mm]  
{\cal I}_J^{[\frac1r ph]}(q) & = & \int {\rm d}r \, r \,
j_J(qr) \, R_p^*(r) \, R_h(r) \, .
\end{eqnarray}

\newpage 

\section*{Appendix B. Two-particle emission matrix elements}

In this appendix we present the expressions of  the matrix elements
involved in the 
two-nucleon emission part of the longitudinal and transverse
responses. The charge operator contributes to the first one and the
corresponding Coulomb matrix elements are:
\begin{eqnarray}
 M^{(\rm 2p2h)}_{2.1} & = &  G_E^{\pon}(q^2) \,
  \delta_{t_{\pon},t_{\hon}} \, \delta_{t_{\ptw},t_{\htw}} \, \sqrt{ 4\pi}
\, (-1)^J \, {\widehat{J}_p} \, {\widehat{J}_h} \, {\widehat{J}} \,
\sum_{LK} \, \frac{\widehat{K}}{\widehat{L}}  \,
\gamma^K_{\pon \hon} \, \gamma^L_{\ptw \htw} \\ \nonumber
&~& 
\ninej{j_{\pon}}{j_{\ptw}}{J_p}{j_{\hon}}{j_{\htw}}{J_h}{K}{L}{J} 
\, \threej{J}{K}{L}{0}{0}{0} \, 
{\cal I}_{JL}^{[\pon \hon];[\ptw \htw]}(q)  \, ,\\[.5cm]
M^{(\rm 2p2h)}_{3.1} & = &  G_E^{\pon}(q^2) \, \delta_{t_{\pon},t_{\hon}} \, 
\delta_{t_{\ptw},t_{\htw}} \, 4\pi \, (-1)^{J+j_{\pon}+j_{\hon}+1} \, 
{\widehat{J}_p} \, {\widehat{J}_h} \, \sum_{iL} \, 
\delta_{t_{i},t_{\pon}} \, \frac {1}{\widehat{L}^2} \\ \nonumber
&~&    \gamma^J_{\pon i} \, \gamma^L_{\ptw \htw} \, \gamma^L_{i \hon} \,  
  \sixj {j_{\ptw}}{j_{\htw}}{L}{j_{\hon}}{j_{i}}{J_h}\,
  \sixj {J_p}{J_h}{J}{j_{i}}{j_{\pon}}{j_{\ptw}}\,
    {\cal I}_{J}^{[\pon i]}(q) \,
    {\cal J}_{L}^{[\ptw \htw];[i \hon]} \, , \\[.5cm]
M^{(\rm 2p2h)}_{3.5} & = &  G_E^{\ptw}(q^2) \, \delta_{t_{\pon},t_{\hon}} \, 
\delta_{t_{\ptw},t_{\htw}} \, 4\pi \, (-1)^{J+j_{\ptw}+j_{\htw}+1} \,
{\widehat{J}_p} \, {\widehat{J}_h} \, \sum_{iL} \,
\delta_{t_{i},t_{\ptw}} \, \frac {1}{\widehat{L}^2} \\ \nonumber
&~&    \gamma^J_{i \htw} \, \gamma^L_{\ptw i} \, \gamma^L_{\pon \hon} \,
  \sixj {j_{\hon}}{j_{\pon}}{L}{j_{\ptw}}{j_{i}}{J_p} \,
  \sixj {J_h}{J_p}{J}{j_{i}}{j_{\htw}}{j_{\hon}} \,
    {\cal I}_{J}^{[i \htw]}(q) \,
    {\cal J}_{L}^{[\pon \hon];[\ptw i]} \, , \\[.5cm]
M^{(\rm 2p2h)}_{3.9} & = &  G_E^{\pon}(q^2) \, \delta_{t_{\pon},t_{\htw}} \,
\delta_{t_{\ptw},t_{\hon}} \, \delta_{j_{\ptw},j_{\hon}} \, 4\pi \,
(-1)^{J+j_{\pon}+j_{\hon}} \, {\widehat{J}_p} \, {\widehat{J}_h} \,
\frac{1}{\widehat{j_{\hon}}^2} \,  \gamma^J_{\pon \htw} \\ &~& \nonumber  
\sixj {J_h}{J}{J_p}{j_{\pon}}{j_{\ptw}}{j_{\htw}} \,
{\cal I}_{J}^{[\pon \htw]}(q) 
\, \sum_{iL} \, \delta_{t_i,t_{\ptw}} \, (-1)^{j_{i}+j_{\htw}} 
\,\frac {1}{\widehat{L}^2} \, \gamma^L_{i \hon} \, \gamma^L_{\ptw i} \,
{\cal J}_{L}^{[\ptw i];[i \hon]} \, , \\[.5cm]
M^{(\rm 2p2h)}_{3.11} & = & G_E^{\pon}(q^2) \, \delta_{t_{\pon},t_{\htw}} \,
\delta_{t_{\ptw},t_{\hon}} \, \delta_{j_{\ptw},j_{\hon}} \, 4\pi \,
(-1)^{J+j_{\pon}+j_{\htw}+1} \, \xi(l_{\ptw}+l_{\hon}) \, 
{\widehat{J}_p}{\widehat{J}_h} 
\\ \nonumber &~& \gamma^J_{\pon \htw} \,
\sixj {J_h}{J}{J_p}{j_{\pon}}{j_{\hon}}{j_{\htw}} \,
{\cal I}_{J}^{[\pon \htw]}(q) \,
{\cal J}_{0}^{[\ptw \hon];[\rho]} \, .
\end{eqnarray}

As already stated, in our model, the convection current is considered
only in 
the uncorrelated diagrams which do not appear in this case. As a
consequence,  only the magnetization current contributes to the 
electric and magnetic transverse terms. For the electric ones, 
the corresponding matrix elements can be written as follows:
\begin{eqnarray}
T^{{\rm E}(\rm 2p2h)}_{2.1;m} & = & q \, \frac{G_M^{\pon}(q^2)}{2M_{\pon}} \,
\delta_{t_{\pon},t_{\hon}} \, \delta_{t_{\ptw},t_{\htw}} \, \sqrt{12} \, 
(-1)^{J+l_{\hon}} \, {\widehat{J}_p} \, {\widehat{J}_h} \, {\widehat{J}}
\, {\widehat{j}_{\pon}} \, {\widehat{j}_{\hon}} \,
{\widehat{l}_{\pon}} \, {\widehat{l}_{\hon}} \\ \nonumber
&~& \sum_{L} \, \frac{1}{\widehat{L}} \,
\gamma^L_{\ptw \htw} \, {\cal I}_{JL}^{[\pon \hon];[\ptw \htw]}(q) \,
\sum_{K} \, {\widehat{K}^2} \, \threej {K}{J}{L}{1}{-1}{0} \,
\ninej{j_{\pon}}{j_{\ptw}}{J_p}{j_{\hon}}{j_{\htw}}{J_h}{K}{L}{J} \\
\nonumber &&
\sum_{\lambda} \, \widehat{\lambda}^2 \, \xi(L+J+\lambda) \, 
\threej{l_{\pon}}{\lambda}{l_{\hon}}{0}{0}{0} \,
\threej{1}{K}{\lambda}{1}{-1}{0} \, 
\ninej{l_{\pon}}{\half}{j_{\pon}}
{l_{\hon}}{\half}{j_{\hon}}
{\lambda}{1}{K} \, , \\[.5cm]
T^{{\rm E}(\rm 2p2h)}_{3.1;m} & = & q \, \frac{G_M^{\pon}(q^2)}{2M_{\pon}} \,
\delta_{t_{\pon},t_{\hon}} \, \delta_{t_{\ptw},t_{\htw}} \, \sqrt{4\pi}
\, (-)^{J+j_{\pon}+j_{\hon}+l_{\pon}+1} \, {\widehat{J}_p} \, 
{\widehat{J}_h}\, {\widehat{J}} \, {\widehat{j}_{\pon}} \\ && \nonumber 
\sum_{iL} \, \delta_{t_i,t_{\pon}} \, \xi(l_{\pon}+l_i+J) \, 
\frac{\widehat{j}_i}{\widehat{L}^2} \, 
\gamma^L_{i \hon} \, \gamma^L_{\ptw \htw} \, {\cal I}_{J}^{[\pon i]}(q)
{\cal J}_{L}^{[\ptw \htw];[i \hon]} \\ && \nonumber
    \sixj {J_p}{J_h}{J}{j_i}{j_{\pon}}{j_{\ptw}} \,
    \sixj {j_{\ptw}}{j_{\htw}}{L}{j_{\hon}}{j_i}{J_h} \,
    \threej {j_i}{j_{\pon}}{J}{-\half}{-\half}{1} \, , \\[5mm]
T^{{\rm E}(\rm 2p2h)}_{3.5;m} & = &  q \, \frac{G_M^{\ptw}(q^2)}{2M_{\ptw}} \,
\delta_{t_{\pon},t_{\hon}} \, \delta_{t_{\ptw},t_{\htw}} \, \sqrt{4 \pi} 
\, (-1)^{J+j_{\ptw}+j_{\htw}+1} \, {\widehat{J}_p} \, {\widehat{J}_h}
\, {\widehat{J}} \, {\widehat{j}_{\htw}} \\ \nonumber &~& 
\sum_{iL} \, \delta_{t_i,t_{\ptw}} \, (-1)^{l_i} 
   \, \xi(l_i+l_{\htw}+J) \,
\frac{\widehat{j}_i}{\widehat{L}^2} 
\, \gamma^L_{\ptw i} \, \gamma^L_{\pon \hon} \, {\cal I}_{J}^{[i \htw]}(q)
\, {\cal J}_{L}^{[\ptw i];[\pon \hon]} \\ \nonumber &&
    \sixj {J_h}{J_p}{J}{j_i}{j_{\htw}}{j_{\hon}} \,
    \sixj {j_{\hon}}{j_{\pon}}{L}{j_{\ptw}}{j_i}{J_p} \,
    \threej {j_{\htw}}{j_i}{J}{-\half}{-\half}{1} \, ,
\\[5mm]
T^{{\rm E}(\rm 2p2h)}_{3.9;m} & = &  q \, \frac{G_M^{\pon}(q^2)}{2M_{\pon}} \,
\delta_{t_{\pon},t_{\htw}} \, \delta_{t_{\ptw},t_{\hon}} \, 
\delta_{{j}_{\ptw},{j}_{\hon}} \, \sqrt{4\pi} \,
   (-1)^{J+j_{\pon}+j_{\hon}+l_{\pon}}
\\ &~& \nonumber 
\xi(l_{\pon}+l_{\htw}+J) \,
  {\widehat{J}_p}\, {\widehat{J}_h}\, {\widehat{J}} \,
  {\widehat{j}_{\pon}}\, \frac{\widehat{j}_{\htw}}{\widehat{{j}_{\ptw}}^2}  
    \sixj {J_h}{J}{J_p}{j_{\pon}}{j_{\ptw}}{j_{\htw}} \,
    \threej {j_{\htw}}{j_{\pon}}{J}{-\half}{-\half}{1} \,
  {\cal I}_{J}^{[\pon \htw]}(q) \\ \nonumber &&
    \sum_{iL} \, \delta_{t_i,t_{\ptw}} \, (-1)^{j_i+j_{\htw}} \, 
    \frac {1}{\widehat{L}^2} \, \gamma^L_{i \hon} \, \gamma^L_{\ptw i}
\, {\cal J}_{L}^{[\ptw i];[i \hon]} \, ,\\[.5cm]
T^{{\rm E}(\rm 2p2h)}_{3.11;m} & = &  q \, 
\frac{G_M^{\pon}(q^2)}{2M_{\pon}} \,
  \delta_{t_{\pon},t_{\htw}} \, \delta_{t_{\ptw}t_{\hon}} \,
  \delta_{j_{\ptw},j_{\hon}} \, \sqrt{4 \pi} \,
   (-1)^{J+j_{\pon}+j_{\htw}+l_{\pon}+1} \\ \nonumber &&
 \xi(l_{\pon}+l_{\htw}+J) \,
\xi(l_{\ptw}+l_{\hon}) \, {\widehat{J}_p}\, {\widehat{J}_h}\, {\widehat{J}}
  {\widehat{j}_{\pon}} \, {\widehat{j}_{\htw}}  \\ \nonumber &&    
    \sixj {J_h}{J}{J_p}{j_{\pon}}{j_{\hon}}{j_{\htw}} \,
    \threej {j_{\htw}}{j_{\pon}}{J}{-\half}{-\half}{1} \,
  {\cal I}_{J}^{[\pon \htw]}(q) \, {\cal J}_{0}^{[\ptw \hon];[\rho]} 
\, .  
\end{eqnarray}

Finally, the magnetic transverse matrix elements of the 
magnetization current are:
\begin{eqnarray}
T^{{\rm M}(\rm 2p2h)}_{2.1;m} & = &  i q \, 
\frac{G_M^{\pon}(q^2)}{2M_{\pon}} \,  
\delta_{t_{\pon},t_{\hon}} \, \delta_{t_{\ptw},t_{\htw}} \, \sqrt{6} \,
(-1)^{l_{\pon}} \, {\widehat{J}_p} \, {\widehat{J}_h} \,
{\widehat{j}_{\pon}} \, {\widehat{j}_{\hon}}
{\widehat{l}_{\pon}} \, {\widehat{l}_{\hon}}
\\ \nonumber
&~&
\sum_{LK} \, (-1)^{K+L} \, \frac{{\widehat{K}^2}}{{\widehat{L}}} \,
\gamma^L_{\ptw \htw} \, 
\ninej{j_{\pon}}{j_{\ptw}}{J_p}{j_{\hon}}{j_{\htw}}{J_h}{K}{L}{J} 
\, \sum_s \, s \, \sqrt{J+\delta_{s,-1}} \, {\widehat{J'}} \,
   {\cal I}_{J' L}^{[\pon \hon];[\ptw \htw]}(q)
\\ \nonumber &~&
\sum_\lambda {\widehat{\lambda}^2}
       \ninej  {l_{\pon}}{\half}{j_{\pon}}
               {l_{\hon}}{\half}{j_{\hon}}
               {\lambda}{1}{K}
       \threej {l_{\pon}}{\lambda}{l_{\hon}}{0}{0}{0}
       \sixj   {L}{J'}{\lambda}{1}{K}{J}
       \threej {L}{J'}{\lambda}{0}{0}{0} \, ,
 \\[.5cm]
T^{{\rm M}(\rm 2p2h)}_{3.1;m} & = & i q \, 
\frac{G_M^{\pon}(q^2)}{2M_{\pon}} \,  
\delta_{t_{\pon},t_{\hon}} \, \delta_{t_{\ptw},t_{\htw}} \, 4\pi \,
(-1)^{j_{\pon}+j_{\hon}+J} \, 
\frac{{\widehat{J}_p}{\widehat{J}_h}}{{\widehat{J}^2}} 
\, \sum_{iL} \, \delta_{t_i,t_{\pon}} \, \frac{1}{{\widehat{L}^2}} 
\, \eta^J_{\pon i} \,
\\ \nonumber
&~& \sixj {j_{\ptw}}{j_{\htw}}{L}{j_{\hon}}{j_i}{J_h} \,
 \sixj {J_p}{J_h}{J}{j_i}{j_{\pon}}{j_{\ptw}} \,
 \gamma^L_{i \hon} \,  \gamma^L_{\ptw \htw} \,
{\cal J}_{L}^{[i \hon];[\ptw \htw]} \\ \nonumber &&
\sum_s \, s \, \sqrt{\frac{J+\delta_{s,-1}}{J+\delta_{s,1}}} \,
(\chi_{\pon}+\chi_i+sJ+\delta_{s,1}) \,  {\cal I}_{J'}^{[\pon i]}(q) 
\, , \\[.5cm]
T^{{\rm M}(\rm 2p2h)}_{3.5;m} & = & i q \, 
\frac{G_M^{\htw}(q^2)}{2M_{\htw}} \, 
\delta_{t_{\pon},t_{\hon}} \, \delta_{t_{\ptw},t_{\htw}} \, 4\pi \,
(-1)^{j_{\htw}+j_{\ptw}+J} \, 
\frac{{\widehat{J}_p}{\widehat{J}_h}}{{\widehat{J}^2}} \, 
\sum_{iL} \, \delta_{t_i,t_{\ptw}} \, \frac{1}{{\widehat{L}^2}} \,
\eta^J_{i \htw} \\ \nonumber &&
 \sixj {j_{\hon}}{j_{\pon}}{L} {j_{\ptw}}{j_i}{J_p}\,
 \sixj {J_h}{J_p}{J}{j_i}{j_{\htw}}{j_{\hon}}\,
 \gamma^L_{\pon \hon} \gamma^L_{\ptw i}\,
{\cal J}_{L}^{[\ptw i];[\pon \hon]} \\ \nonumber &&
\sum_s \,s \,\sqrt{\frac{J+\delta_{s,-1}}{J+\delta_{s,1}}} \,
(\chi_{\htw}+\chi_i+sJ+\delta_{s,1})\, {\cal I}_{J'}^{[i \htw]}(q)
\, , \\[.5cm]
T^{{\rm M}(\rm 2p2h)}_{3.9;m} & = & -i q \, 
\frac{G_M^{\pon}(q^2)}{2M_{\pon}}\, 
\delta_{t_{\pon},t_{\htw}} \, \delta_{t_{\ptw},t_{\hon}} \, 
\delta_{{j}_{\hon},{j}_{\ptw}} \, 4 \pi \,
(-1)^{J+j_{\pon}+j_{\hon}} \, 
\frac{{\widehat{J}_p}{\widehat{J}_h}}{{\widehat{J}^2}
{\widehat{{j}_{\hon}}^2}} \, \eta^J_{\pon \htw} \,
\\ \nonumber
&~&
 \sixj{J_h}{J}{J_p}{j_{\pon}}{j_{\ptw}}{j_{\htw}} \,
 \sum_s \, s \, \sqrt{\frac{J+\delta_{s,-1}}{J+\delta_{s,1}}} \,
 (\chi_{\pon}+\chi_{\htw}+sJ+\delta_{s,1})\,
 {\cal I}_{J'}^{[\pon \htw]}(q) \\ \nonumber
&~& \sum_{iL} \, \delta_{t_i,t_{\ptw}} \,
 (-1)^{j_i+j_{\htw}}\, \frac{1}{{\widehat{L}^2}}\,
 \gamma^L_{i \hon} \, \gamma^L_{\ptw i}\,
{\cal J}_{L}^{[\ptw i];[i \hon]} \, , \\[.5cm]
T^{{\rm M}(\rm 2p2h)}_{3.11;m} & = & i q \, 
\frac{G_M^{\pon}(q^2)}{2M_{\pon}} \,
\delta_{t_{\pon},t_{\htw}} \, \delta_{t_{\ptw},t_{\hon}} \,
\delta_{j_{\ptw},j_{\hon}} \, 4\pi \,
(-1)^{J+j_{\pon}+j_{\htw}} \, \xi(l_{\ptw}+l_{\hon}) \,
\frac{{\widehat{J}_p}{\widehat{J}_h}}{{\widehat{J}^2}} \, 
\eta^J_{\pon \htw} 
\\ \nonumber
&~&
\sixj {J_h}{J}{J_p} {j_{\pon}}{j_{\hon}}{j_{\htw}} \,
{\cal J}_{0}^{[\ptw \hon];[\rho]} \,
\sum_s \, s \, \sqrt{\frac{J+\delta_{s,-1}}{J+\delta_{s,1}}} \,
(\chi_{\pon} +\chi_{\htw}+sJ+\delta_{s,1})\,
{\cal I}_{J'}^{[\pon \htw]}(q) \, .
\end{eqnarray}

The integrals and symbols used in these equations are defined in
Appendix A.

\end{document}